\documentclass[sigconf,natbib=true]{acmart}
\usepackage{bm}
\usepackage{algorithm}
\usepackage{algorithmic}
\usepackage{comment}
\usepackage{microtype}
\usepackage{subfigure}
\usepackage{graphicx}
\usepackage{multirow}
\usepackage{enumitem}
\usepackage{titlesec}
\AtBeginDocument{%
	\providecommand\BibTeX{{%
			\normalfont B\kern-0.5em{\scshape i\kern-0.25em b}\kern-0.8em\TeX}}}

\setcopyright{acmcopyright}
\copyrightyear{2022}
\acmYear{2022}
\acmDOI{10.1145/3477495.3532074}
\acmConference[SIGIR '22]{The 45th International ACM SIGIR Conference on Research and Development in Information Retrieval}{11-15 July 2022}{Madrid, Spain}
\acmBooktitle{The 45th International ACM SIGIR Conference on Research and Development in Information Retrieval, 11-15 July 2022, Madrid, Spain}
\acmPrice{15.00}
\acmISBN{978-1-4503-8732-3/22/07}

\begin{document}
	
\title[User-Centric Conversational Recommendation with Multi-Aspect User Modeling]{User-Centric Conversational Recommendation\\ with Multi-Aspect User Modeling}

\author{Shuokai Li$^{1,2,*}$, Ruobing Xie$^{3,*}$, Yongchun Zhu$^{1,2}$, Xiang Ao$^{1,2,\dag}$, Fuzhen Zhuang$^{4,5}$, Qing He$^{1,2,\S}$}
\affiliation{%
 \institution{$^1$Key Lab of Intelligent Information Processing of Chinese Academy of Sciences (CAS), Institute of Computing \\Technology, CAS, Beijing 100190, China.
 $^2$University of Chinese Academy of Sciences, Beijing 100049, China.\\
 $^3$WeChat Search Application Department, Tencent, China.
 $^4$Institute of Artificial Intelligence, Beihang\\ Unversity, Beijing 100191, China. $^5$SKLSDE, School of Computer Science, Beihang University, Beijing 100191, China.\\
 \{lishuokai18z, zhuyongchun18s, aoxiang, heqing\}@ict.ac.cn,
 ruobingxie@tencent.com,
 zhuangfuzhen@buaa.edu.cn}\country{}}
\thanks{* Shuokai Li and Ruobing Xie have equal contributions.}
\thanks{$\dag$ Xiang Ao is also at Institute of Intelligent Computing Technology, Suzhou, China.}
\thanks{$\S$ Qing He is the corresponding author.}

\renewcommand{\shortauthors}{S. Li et al.}

\begin{abstract}
Conversational recommender systems (CRS) aim to provide high-quality recommendations in conversations. However, most conventional CRS models mainly focus on the dialogue understanding of the current session, ignoring other rich multi-aspect information of the central subjects (i.e., users) in recommendation. In this work, we highlight that the user's historical dialogue sessions and look-alike users are essential sources of user preferences besides the current dialogue session in CRS. To systematically model the multi-aspect information, we propose a User-Centric Conversational Recommendation (UCCR) model, which returns to the essence of user preference learning in CRS tasks. Specifically, we propose a historical session learner to capture users' multi-view preferences from knowledge, semantic, and consuming views as supplements to the current preference signals. A multi-view preference mapper is conducted to learn the intrinsic correlations among different views in current and historical sessions via self-supervised objectives. We also design a temporal look-alike user selector to understand users via their similar users. The learned multi-aspect multi-view user preferences are then used for the recommendation and dialogue generation. In experiments, we conduct comprehensive evaluations on both Chinese and English CRS datasets. The significant improvements over competitive models in both recommendation and dialogue generation verify the superiority of UCCR. The source code will be available on \href{https://github.com/lisk123/UCCR}{https://github.com/lisk123/UCCR}.
\end{abstract}





\maketitle

{
\fontsize{8pt}{8pt} 
\selectfont
\textbf{ACM Reference Format:}
\\
Shuokai Li, Ruobing Xie, Yongchun Zhu, Xiang Ao, Fuzhen Zhuang, Qing He. 2022. User-Centric Conversational Recommendation with Multi-Aspect User Modeling. In  \textit{Proceedings of the 45th International ACM SIGIR Conference on Research and Development in Information Retrieval, 11-15 July 2022, Madrid, Spain} ACM, New York, NY, USA, 10 pages. https://doi.org/10.1145/3477495.35\\32074 }

\section{Introduction}

In recent years, great efforts have been made to develop the conversational recommender systems (CRS) \cite{lesi2020interactive,zhou2020improving,xu2021adapting,fu2021hoops,ren2021learning,xie2021comparison}, which aim to provide high-quality recommendations for users in conversations. Generally, CRS methods can be roughly divided into a recommender module and a dialogue module \cite{chen2019towards,zhou2020improving}. The \emph{dialogue module} converses with users through natural language. In contrast, the \emph{recommender module} learns user preferences based on the dialogue contents, and provides appropriate recommendations for users. For the generative CRS~\cite{zhou2020improving,lu2021revcore}, the recommended items are naturally integrated into the natural language replies and given to users.
Different from traditional recommender systems \cite{cheng2016wide}, CRS mainly captures user preferences according to the current dialogue session, and thus should handle both natural language understanding and user modeling \cite{jannach2020survey}.
Currently, it has also been widely used in various real-world scenarios, such as intelligent voice assistants (e.g., ``Siri'') and customer services on E-commerce platforms.

%1、简介CRS任务，生成式CRS：对话部分理解+生成，核心目的和形式，重要性

To model users' preferences and provide high-quality recommendations, lots of CRS models focus on better natural language understandings. Some works enhance the dialogue representation learning with more sophisticated encoders \cite{li2018towards,zhou2020improving}. Other methods also introduce useful external information such as knowledge graphs (KGs) \cite{chen2019towards,zhou2020improving} and user reviews \cite{lu2021revcore}.
However, most of them pay too much attention to the current dialogue session and only learn the preferences reflected by the session (although we admit that it is indeed an important source of user preferences), ignoring the central subjects in CRS, i.e., \textbf{users}. In practical systems, users usually have various multi-aspect features such as user's historical dialogue sessions and user profiles besides the current session, which could help to provide more comprehensive understandings of users from different perspectives. However, there is seldom work that focuses on user-centric preference learning in CRS.

\begin{figure}[!htpb]%%图
	\centering  %插入的图片居中表示
	\vspace{0mm}
	\includegraphics[width=0.40\textwidth]{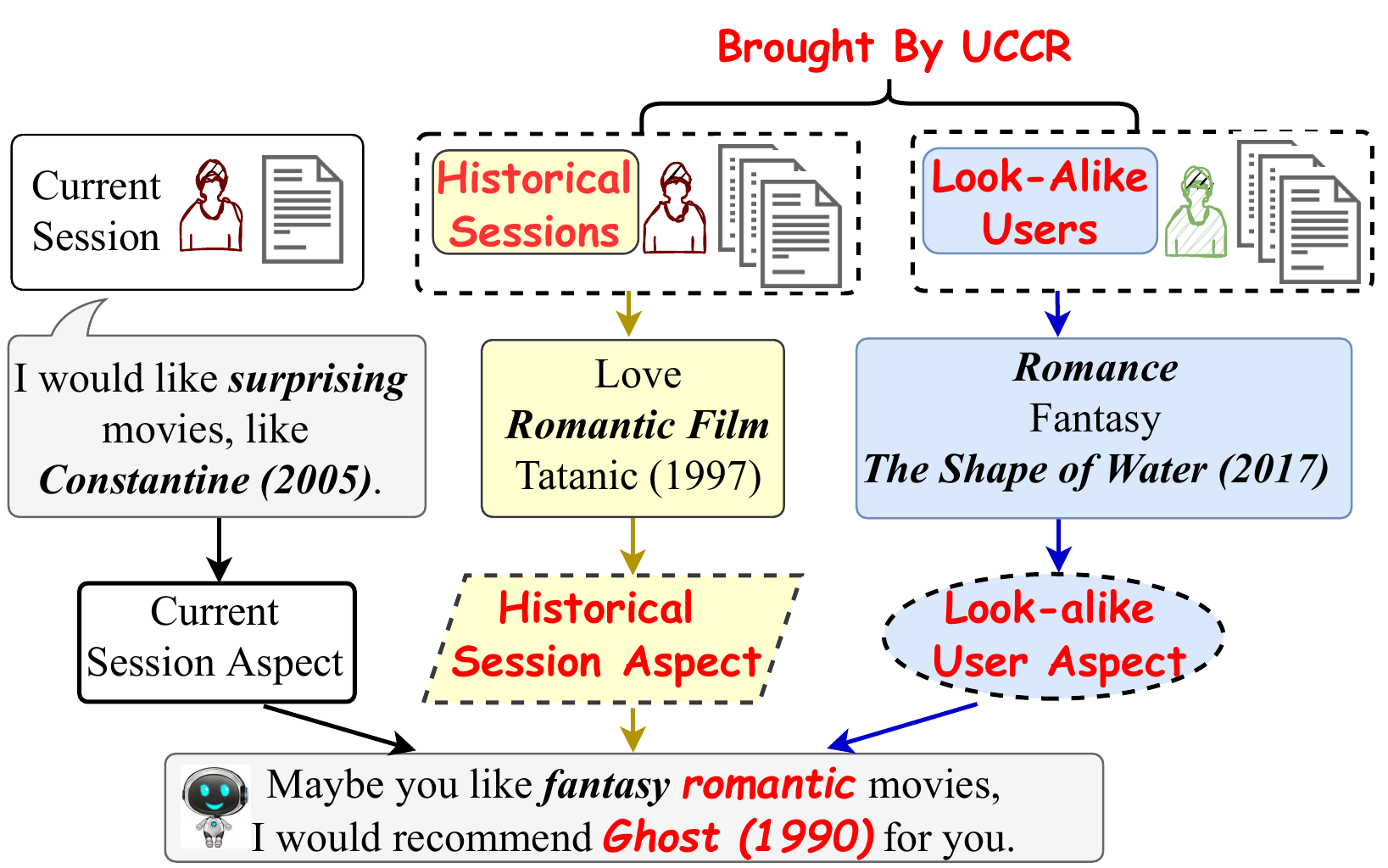}
	\caption{An example of the multi-aspect user information. UCCR introduces the historical dialogue sessions and look-alike users to CRS for user-centric preference learning.}
\label{framework}
	\vspace{-2mm}
\end{figure}

%2、为了核心目的推荐+用户理解，很多CRS工作着重于对话理解
%1. 提升dialogue建模
%2. 引入额外的KG和用户评论知识
%但是他们很少考虑推荐这一侧的信息——最基本的属性是人
%人和session的不同：1、人的信息量更多，2、人和session的粒度不一样（一个人有不同的对话）

% From the current session, we can see the target user is interested in surprising movies. However, from historical data, he also likes romantic movies. Thus we should recommend fantasy romantic movies to him. The look-alike users also confirm his preferences.

% 【】图上端感觉有点乱，look-alike建议加上色块，这里的核心突出点是三个信息源
% 【】图上是否可以强调后两者是我们引入的？并且更宽些
% 【】related work上强调user-centric和current-session-centric的差别

% 图：文档尽量用一样的
% 3个信息源，current是已有工作用的。current用实线圈起来，另外两个用红线框起来
% 下面写全程，historical session aspect 
% 模型复杂：符号体系，不够精炼，原因相关
% related work强调和传统推荐的区别
% user 

In this work, we attempt to emphasize users and polish the model's ability on user-centric preference learning in CRS. As in Fig.~\ref{framework}, the user preferences in real-world CRS could be mainly extracted from three aspects:
(1) the user's \textbf{current dialogue session}, which is the main information widely adopted by conventional CRS models.
%Multi-view textual features (e.g., words in dialogues) and structural features (e.g., entities in KGs) can be used in dialogue understanding.
(2) The user's \textbf{historical dialogue sessions}, which stores user's historical preferences from multiple views.
%the entity-level knowledge view, the word-level semantic view, and the item-level consuming view. 
This historical information is beneficial since users tend to have similar preferences with their historical behaviors, which is inspired by the idea of item-CF \cite{sarwar2001item}.
(3) The user's \textbf{look-alike users}, which could be retrieved by the relevance of user profiles or user historical behaviors. It learns users' preferences via their similar users under the instruction of user-CF~\cite{zhao2010user}.
The newly-introduced information on historical dialogue sessions and look-alike users is beneficial especially when the current session contains little information.

%3、我们的目的是引入人的建模
%然后提一下人可以从多方面信息中得知其偏好：
%1、当前session （已有工作，当前工作都在model，确实是最重要的）（主动获取信息）
%2、历史session  （很少有人做这个）（被动获取信息，基于相似item在未来有用的逻辑）
%3、相似的人，在session信息稀缺的情况下，可以通过userCF扩大推荐可能（当前用户当前、历史不足）（和推荐不同：session，其他信息不足时，相似用户可以补充）（被动获取信息）
%其中23两点都是已有工作基本没有考虑到的，没有联合考虑
%（具体样例可以参考fig，不过感觉可以精简）（fig-1定义好历史session）

However, incorporating multi-aspect user information in CRS is non-trivial, since it is challenging to decide how much we should learn from historical and look-alike features without confusing the current session modeling. Differing from users in classical recommender systems, users in CRS will actively interact with the system via natural language. Hence, their user intentions are more explicit and definite according to the current sessions, and thus the historical and look-alike features should be considered under the constraints of the current user intentions. We hope to smartly utilize multi-aspect features, successfully capturing both the basic \emph{fantasy} intention from the current session and the hidden \emph{romantic} preference from the historical and look-alike features in Fig. \ref{framework}.

% 结合图1的例子

\textls[-0]{In light of the observations above, we propose a novel \textbf{User-Centric Conversational Recommendation (UCCR)} framework to jointly model user's multi-aspect information in CRS.
Specifically, UCCR learns the multi-aspect user preferences mainly from three information sources, including the user's current dialogue session, historical dialogues sessions, and look-alike users. UCCR mainly consists of four parts:
(1) We first design a historical session learner to capture users' diverse preferences in their historical sessions besides learning from the current session. Precisely, we extract multi-view user preferences from the dialogues, including the word-level semantic view, entity-level knowledge view, and item-level consuming view. The correlations between the current and historical information are also considered in historical preference learning.
(2) We propose a multi-view preference mapper to learn the intrinsic correlations among different views in the current/historical sessions. The main idea is that two views of a user should be more relevant, since they reflect similar preferences of the same user. We design three self-supervised cross-view objectives between these views as supplements to the supervised losses, which enables a more sufficient training of user multi-view preferences.
(3) For the look-alike user aspect, we refer to the similar users' preferences as a user-CF based supplement to the target user's understanding. User basic profiles and user historical behaviors, which are essential sources of personalization, could be used for the user similarity calculation. A temporal look-alike user selector is designed for more precise user generalization.
(4) Finally, multi-aspect user-centric modeling is conducted to jointly encode multi-aspect multi-view user preferences into the final user representation.}

\textls[-20]{Through UCCR, these multi-aspect features are properly incorporated under the guiding ideology of user-centric modeling. Compared with conventional CRS models that focus on current session understanding, our UCCR comprehensively understands users from multiple aspects (current dialogue session, historical dialogue sessions, and look-alike users) and multiple views (word, entity, and item views), which returns to the essence of user understanding in recommendations. We summarize the contributions of this work as follows:}
%4、提出模型解决这个问题
%User-Centric Conversational Recommendation with Multi-aspect (Multi-view) User Modeling
%首先是Multi-aspect，包括以上三点
%然后对于session的建模（更细致的model），我们使用multi-view
%在multi-view之间建立CL任务联系，加大不同view之间信息交互
%最后进行prediction
%In experiments, we conduct extensive evaluations on both the dialogue and recommendation parts in two real-world CRS datasets. The significant improvements over the SOTA baselines and ablation studies demonstrate the effectiveness and robustness of our proposed UCCR model. 
%Moreover, we also conduct several ablation tests and parameter analyses with detailed discussions to enable a deeper understanding of our model as well as the personalization in CRS. 
\begin{itemize}[leftmargin=*]
  \item We emphasize the user-centric modeling in CRS, and systematically highlight and verify the significance of historical dialogue sessions and look-alike users, returning to the essence of user understanding in CRS. To the best of our knowledge, we are the first to jointly model current dialogue session, historical dialogue sessions, and look-alike users via a user-centric manner in CRS.
  \item We propose a set of techniques to precisely extract useful user preferences related to the current user intentions from multiple views, including a historical session learner, a multi-view preference mapper, and a temporal look-alike user selector.
  \item UCCR achieves the best performances against SOTA baselines on both the dialogue and recommendation parts in two real-world datasets. Extensive model analyses and ablation tests also help to better understand multi-aspects user information in CRS, shedding light on real-world applications.
  %\item UCCR achieves the best performances against SOTA baselines on all metrics. Extensive model analyses and ablation tests also help to better understand multi-aspects user information in CRS, shedding light on real-world applications.
\end{itemize}

% 【】目前novelty太简单，看起来和已有工作太接近导致显得没有创新点

\section{Preliminaries}

% 【】模型的复杂度
% 【】review对CRS的基本模型认知不足，需要介绍一下，比如entity、GCN、DBPedia等这些
% 简单提relation R-GCN
% background of CRS: item都是entity
% 统一定义：entity word item定义比较好。session turn的概念。一个session有多个turn。把每个词的定义完全对齐。介绍方法的时候，尽量别说conversation，说的更准确一些。

\noindent
\textbf{Background of CRS.}
Modern CRS methods~\cite{chen2019towards} aim to provide high-quality items through a multi-turn dialogue with users. Thus they consist of two major components, namely {recommendation module} and {dialogue generation module} \cite{zhou2020towards}.
The \emph{dialogue generation module} aims to generate utterances and converse with users. Each dialogue session may consist of multiple turns. In each dialogue \emph{turn}, the dialogue module could either interact with users or recommend an item. The \emph{recommendation module} aims to provide proper items according to the information in sessions.

There are mainly three objects in CRS, namely user mentioned entities, words, and items. \emph{User mentioned entities} $e\in\mathcal{E}$ are entities in certain KG extracted from dialogues, which contain structural knowledge. In contrast, \emph{words} $w\in\mathcal{W}$ reflect semantic knowledge in dialogues.
%Following~\cite{chen2019towards,zhou2020improving}, user mentioned words and entities are extracted by entity linking~\cite{daiber2013improving}.
In conventional CRS, \emph{items} $d\in\mathcal{I}$ are recommended mainly via user preferences learned from user mentioned entities and words in the current dialogue sessions \cite{zhou2020improving}. Note that in our setting, all items (e.g., movies) are also entities in $\mathcal{E}$ \cite{li2018towards,zhou2020towards}.

\noindent
\textbf{Notions of our UCCR.}
In real-world CRS, a user may have multiple dialogue sessions with the system. We organize user dialogue sessions in chronological order. For a user $u$ from $\mathcal{U}$ having $T$ dialogue sessions, we have the following definitions:

$\textit{\textbf{Definition 1: Current Dialogue Session}}.$ We regard the $T$-th session as the current dialogue session that we should recommend for. In the current (dialogue) session, when recommending items at a certain turn, all $t$ user mentioned entities $\mathcal{C}_e=\{e_1^T,...,e_{t}^T\}$ of the current session before this turn are viewed as the \emph{current entities}. For words, the definition of \emph{current words} $\mathcal{C}_w$ is the same as $\mathcal{C}_e$.

%the current words $\mathcal{C}_w=\{w_1^T,...,w_{t}^T\}$ and current entities $\mathcal{C}_e=\{e_1^T,...,e_{t}^T\}$ are collected. Here
%For the $t$-th recommendation turn, the current dialogue session includes current words $\mathcal{C}_w=\{w_1^T,...,w_{t-1}^T\}$ and current entities $\mathcal{C}_e=\{e_1^T,...,e_{t-1}^T\}$. Here $w_m^j$/$e_m^j$ denotes the words/entities from the $m$-th recommendation turn of $j$-th dialogue~(for the current session, $j=T$).

\textls[-12]{$\textit{\textbf{Definition 2: Historical Dialogue Sessions}}.$ We call all previous sessions before the current session as historical dialogue sessions. It also includes the \emph{historical entities} $\mathcal{H}_e=\{\mathcal{H}_e^1, \cdots, \mathcal{H}_e^{T-1}\}$ and \emph{historical words} $\mathcal{H}_w$ extracted from all $T-1$ sessions. Besides, the previous recommended items in historical sessions are viewed as \emph{historical preferred items} $\mathcal{H}_d$.
%, reflecting user preferences. 
Precisely, $\mathcal{H}_e^{j}=\{e_1^{j},...e^j_{t_j}\}$ includes all $t_j$ user mentioned entities of the $j$-th historical dialogue session, and similar as $\mathcal{H}_w$ and $\mathcal{H}_d$.
We should double clarify that our proposed historical dialogue session is completely different from the dialog/conversation history used in \cite{chen2019towards,zhou2020improving,lu2021revcore}, for their ``historical'' information locates in the historical sentences (turns) of the current dialogue session.
To the best of our knowledge, we are the first to highlight the significance of historical dialogue sessions in generative CRS.}

% 【】需要说明已有很多工作的不一样，以前的只是historical sentence in current sessions，加引用，比如RevCore

$\textit{\textbf{Definition 3 (Look-alike Users)}}.$
The look-alike users refer to similar users. The user similarity can be calculated from multiple perspectives, such as user profiles and historical behaviors. In UCCR, we rely on the historical words, entities, and items for look-alike users learning.
%we could use user profiles, behavioral data,
%For a user $u$, the users who have similar historical preference views~(i.e. word, entity or item) are his look-alike users.
In light of user-CF, the look-alike users may have similar tastes, thus could enhance the user representations, which is especially effective when users only have sparse information learned from the current or historical dialogue sessions.

%Def.2 and 3 are the new information introduced by our work.

\begin{figure*}[htpb]%%图
	\centering  %插入的图片居中表示
	\includegraphics[width=0.92\textwidth]{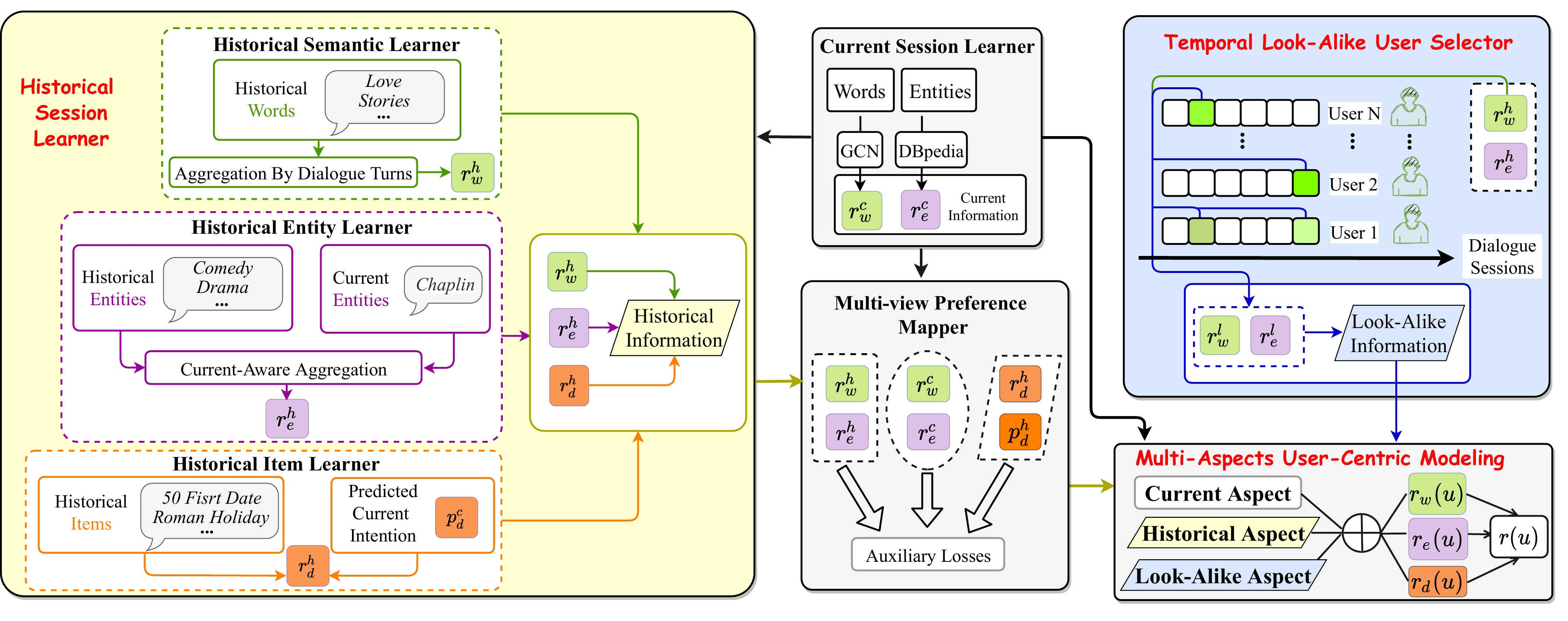}
	\vspace{-0mm}
	\caption{\textls[-12]{The overview of our model UCCR. %We highlight the central object (i.e., users) from multiple aspects.
	First, the multiple views information is encoded by the historical and current session learners. Second, the multi-view preference mapper further explores the correlations between views. Next, the temporal look-alike users selector provides another aspect feature. Finally, these aspects are fused by the user-centric modeling module.}}
\label{network}
	\vspace{-2mm}
\end{figure*}

% 【】空格子需要填上内容
% 【】整个图的思路不太明显，可以使用不同的颜色链接边，突出逻辑路径，突显不同的信息流，另外线条太复杂
% temporal：画个时间轴，很多用户在时间轴
% 三种信息源，三个不同色块、historical的字（new proposed）、边框、流程线，都加颜色
% 图上的符号太多了，只保留r相关的符号
% 三个信息源、三条路径信息清晰、两条路径是新提出、符号少一些
% current-aware influenced

\section{Method}
\label{sec_method}

% 【】模型很复杂？？需要补上各个motivation，比如已有工作就加引用，新模型就加上motivation

In this section, we propose our User-Centric Conversational Recommendation (UCCR) to the CRS task. Unlike conventional CRS methods \cite{chen2019towards,zhou2020improving,lu2021revcore} that merely focus on the current dialogue session, our UCCR jointly models multi-aspect user features including (1) current dialogue session, (2) historical dialogue sessions, and (3) look-alike users for comprehensive user understandings.

The user-centric framework works as follows:
First, in both current and historical sessions, we jointly consider word, entity, and item views to model user current and historical preferences. The historical session learner is specially designed to effectively extract useful information related to the current user intention (Sec. \ref{sec.current_session} and \ref{his_sess_learner}).
Second, we propose the multi-view preference mapper to learn the intrinsic correlations among words, entities, and items in both current and historical sessions via multiple self-supervised objectives (Sec. \ref{multi_view_align}).
For look-alike users, as user preferences are changing dynamically, we also design a temporal look-alike user selector to find more appropriate similar users (Sec. \ref{temp_look_alike_u_sel}).
Finally, the overall user preferences are learned by jointly considering multi-aspect and multi-view user features (Sec. \ref{sec.u_centric_m}).
The overview illustration of our UCCR framework is shown in Fig.~\ref{network}.

\subsection{Current Session Learner}
\label{sec.current_session}

We first introduce how to encode the user features from the current dialogue session~(i.e., current words $\mathcal{C}_w$ and current entities $\mathcal{C}_e$).

\subsubsection{Current Entity Learner}

% 【】这里entity是直接被mark出来的吧？

Following \cite{chen2019towards,zhou2020improving}, we use the widely-used knowledge graph DBpedia~\cite{lehmann2015dbpedia} as the entity source. It stores factual knowledge triples $<e_1,r,e_2>$, where $e_1, e_2\in\mathcal{E}$ are entities, and $r\in\mathcal{R}$ is the relation. Note that the user mentioned entities are pre-marked and fixed via DBpedia in our CRS datasets \cite{li2018towards,zhou2020towards}.
% 【】这里补一下用DBpedia的baseline的引用，以及下面的拥R-GCN的文章的引用
% R-GCN(C_e)=Softmax()(V_e)
As entity relation is important to consider, following \cite{chen2019towards,zhou2020improving}, we adopt the powerful R-GCN~\cite{schlichtkrull2018modeling} to encode the structural triple information into entity representations as follows:
\begin{equation}
\label{rgcn}
    \bm{v}_e^{l+1} = \sigma(\sum_{r\in\mathcal{R}}\sum_{e'\in\mathcal{N}_e^r}\frac{1}{\textrm{Z}_{e,r}}W_r^l\bm{v}_{e'}^l+W^l\bm{v}_e^l),
\end{equation}
% 【】review1提示公式错误。
% 【重要】另外，矩阵和向量需要\bm{}，全文修改
where $\bm{v}_e^l\in\mathbb{R}^d$ is the $l$-th layer's representation of entity $e$, and $\mathcal{N}_e^r$ is the one-hop neighbor set of $e$ under the relation $r$. $W_r^l$ and $W^l$ are trainable weights of layer $l$, and $\textrm{Z}_{e,r}$ is a normalization factor. For convenience, we use the last layer's representation $\bm{v}_e^{L}$ as the entity representation $\bm{v}_e$. Via Eq.~\ref{rgcn}, the current entities $\mathcal{C}_e = \{e_1^T, \cdots, e_t^T\}$ are transformed into an entity matrix $\bm{V}_e=\{\bm{v}_{e_1^T}, \cdots,\bm{v}_{e_{t}^T}\}$.
% 【】有些符号问题我修改了，check一下是否正确
% input: C_w，entity提一下V_e，后面没必要用V_e这个符号了。
Next, following~\cite{chen2019towards,zhou2020improving}, we incorporate the self-attention mechanism to aggregate the entity matrix $\bm{V}_e$ according to the importance of entities. The final current entity representation $\bm{r}_e^c$ is built as:
\begin{equation}
\label{cur_entity}
    \begin{aligned}
    \bm{r}_e^c &= \textrm{R-GCN}(\mathcal{C}_e) =  \mathcal{F}(\bm{\mu}_e(\bm{V}_e)^\textrm{T}),\\
    \bm{\mu}_e &= \textrm{Softmax}(\bm{b}_e\textrm{Tanh}(W_e\bm{V}_e)),
    \end{aligned}
\end{equation}
% lisk: 试了把这两个式子合成一行，但是看起来太乱了，还是两行吧
% 【】符号体系缺失有点乱...比如这里的r表示entity结果？直接e_c可以吗？
% 【】另外这个算是self-attention吗...
where $\mathcal{F}:\mathbb{R}^d\to\mathbb{R}^d$ is a linear transformation, $\bm{b}_e\in\mathbb{R}^{1\times d}$ and $W_e\in\mathbb{R}^{d\times d}$ are trainable weights.
%, and $\bm{r}_e^c\in\mathbb{R}^{1\times d}$ are the representations of current entities.
%Here the current entity embeddings of entities contain the structural-based prior knowledge in external knowledge graph DBpedia.}
%which reflects the user current preferences on entity-level knowledge view.

\subsubsection{Current Semantic Learner}
\label{cur_word_learner}

The structural knowledge in entities reflects user preferences accurately, but lacks generalization. According to~\cite{zhou2020improving}, the semantic information of words could effectively improve the ability of preference generalization.
Following \cite{zhou2020improving}, we adopt an external lexical dataset ConceptNet~\cite{speer2017conceptnet} to bring in prior semantic information. The semantic similarities from the dataset are used to build edges between words.
Precisely, given the words in the current dialogue session $\mathcal{C}_w=\{w_1^T,...,w_t^T\}$, we first leverage GCN to learn the embeddings of current words. Then the current semantic representation $\bm{r}_w^c=\textrm{GCN}(\mathcal{C}_w)$ is calculated via self-attention similar as Eq. (\ref{cur_entity}).
%Precisely, we leverage GCN \cite{edwards2016graph} as the neighbor aggregator to build the word matrix $\bm{V}_w=\{\bm{v}_{w_1^T},...,\bm{v}_{w_{t}^T}\}$ for all current words, where words embeddings $\bm{v}_{w_i^T}$ are learned by GCN. The final current semantic representation $\bm{r}_w^c$ is learned via self-attention similar as Eq. (\ref{cur_entity}).

\subsection{Historical Session Learner}
\label{his_sess_learner}
% 强调hist和current如何链接上的

In this section, we introduce the historical information to improve user preference learning. However, directly calculating historical session information as the current session in Sec. \ref{sec.current_session} will lead to poor performances, since the gap between user's historical and current intentions may confuse the current recommendation. We believe that the historical sessions should better work as supplements under the constraints of users' current intentions, providing additional information to complement user preferences.
Hence, we propose a multi-view historical session learner to capture current-related historical information from entity, semantic, and item views.

\subsubsection{Historical Entity Learner}

For the historical entities $\mathcal{H}_e=\{\mathcal{H}_e^1,...,\mathcal{H}_e^{T-1}\}$, we first learn an entity-level session representation for each dialogue $\mathcal{H}_e^j$, and then aggregate the $T-1$ sessions according to the similarities between historical and current sessions to learn the historical entity representation $\bm{r}_e^h$.

Different from other recommender systems, users in CRS expect high-quality recommendations mainly based on the current session. The current entities are very important to reflect users' current intentions (refer to Sec.~\ref{ablation}).
%As the entity representations consist of fine-grained structural knowledge and models user preferences accurately~(refer to the experiments in Sec.~\ref{ablation}), the current entities are more reliable than historical entities for user modeling.
Hence, we attempt to extract user's entity-view preferences related to the current entities from historical entities in the historical entity learner.
%Thus we pay more attention to current session for entities, and extract relevant entities information from historical entities according to the similarity between historical sessions and current session.
%The historical entity learner has the same idea as word learner: first calculating the average vectors for each session, then averaging them to learn historical entity representation $\bm{r}_e^h$. Different from the semantic words, the entity representations contain structural knowledge as external knowledge graphs are incorporated. Thus the importance of entities mainly depends on the structural information. As the entity representations contain structural knowledge, we use the self-attention of entities to calculate the weights, rather than the dialogue turns.
Specifically, given the $j$-th historical entity set $\mathcal{H}_e^j=\{e_1^j,...,e_t^j\}$, we use the same R-GCN and self-attention in Eq. (\ref{rgcn}) and (\ref{cur_entity}) to learn the $j$-th session's entity representation $\bm{h}_e^j$.
Next, the similarity between these learned entity representations and the current entity $\bm{r}_e^c$ is considered to weigh $T-1$ historical sessions' entity representations. The final historical entity representations $\bm{r}_e^h$ is learned as:
\begin{equation}
\label{his_entity}
    \begin{aligned}
    \bm{r}_e^h &= \textrm{Agg}(\bm{r}_e^c, \bm{h}_e^1,...,\bm{h}_e^{T-1}) = \sum_{j=1}^{T-1} \varphi(\bm{h}_e^j, \bm{r}_e^c)\ \bm{h}_e^j,\\
    \varphi&(\bm{h}_e^j, \bm{r}_e^c) = \textrm{Softmax}(\bm{h}_e^j W_s\bm{r}_e^c\ /\ \lambda_e),
    \end{aligned}
\end{equation}
where $\textrm{Agg}(\cdot)$ denotes the attention-based weighted aggregation of historical entities. Through Eq. (\ref{his_entity}), the current-related historical entity-view information is successfully extracted.
%Note that the difference between CRS and traditional recommendations is that CRS captures user current features, which reflects current user intentions. Thus our Eq.~\ref{his_entity} differs from traditional recommendation tasks and fits for the CRS scenario.

\subsubsection{Historical Semantic Learner}

%The idea of historical semantic learner is the same as entity: first learning one vector for each historical session, then aggregating them to learn historical word representation $\bm{r}_w^h$.

We also attempt to learn historical semantic-view information correlated to the current user intention. Differing from entities, we highlight the temporal factors for modeling historical words. 
The semantic information close to the current session is prone to be more important than previous sessions. Hence,  considering simplicity and efficiency, using temporal factors is enough to well extract useful current-related semantic information from historical words.

Concretely, given the $j$-th historical word set $\mathcal{H}_w^j=\{w_1^j,...,w_t^j\}$, we first use the same GCN to learn the word representations $\bm{V}_w^j=\{\bm{v}_{w_1^j},...,\bm{v}_{w_t^j}\}$ like Section~\ref{cur_word_learner}. Then the $j$-th session representation is calculated as:
\begin{equation}
\label{eq.his_word}
    \begin{aligned}
    \bm{h}_w^j &= \mathcal{F}(\sum_{m=1}^t s({w_m^j})\ \bm{v}_{w_m^j}),
    %s(m) &= \textrm{Softmax}(\{1,2,...,t\})[m],
    \end{aligned}
\end{equation}
% 【】这里多次historical words/entities都写成sessions了，全文需要反复检查校对
where $s({w_m^j})=\textrm{Softmax}({1,2,...,t})[m]$ is the importance of $w_m^j$ according to the dialogue turn. Through this, a larger $m$ (i.e., a latter word $w_m^j$) will have a larger weight, which follows the assumption that the latter words are supposed to have more useful information. Then for each user, his/her all historical session representations are also aggregated by the temporal factors: $\bm{r}_w^h = \sum_{j=1}^{T-1} s(\bm{h}_w^j)\ \bm{h}_w^j$, where $j$ is the index of the dialogue sessions.
%where $i$ is the index of the recommendation turns. A larger $i$ means that $w_i^j$ appears later, thus has a larger value of $s(i)$. For each historical session, we have a vector $\bm{h}_w^j$ and derive the historical word representations as a weighted sum:
\begin{comment}
\begin{equation}
    \begin{aligned}
    \bm{r}_w^h &= \sum_{j=1}^{T-1} s(\bm{h}_w^j)\ \bm{h}_w^j,
    %c(j) &= \textrm{Softmax}(\{1,2,...,T-1\})[j],
    \end{aligned}
\end{equation}
where $j$ is the index of the dialogue sessions. Finally, the historical word representation $\bm{r}_w^h$ reflects user's historical semantic behaviors.
\end{comment}
%For each historical dialogue session, it contains a number of semantic words. However, not all words contribute to the user preference modeling, and some of words are noisy. Thus, for each session, we average the

\subsubsection{Historical Item Learner}

Here we consider the historical items, which is also an important source for user modeling, as the goal of CRS is to predict the user preference on items.
%For the historical items, as the whole item set $\mathcal{I}$ is included in the entity set $\mathcal{E}$~(recall that entity includes item entity and non-item entity), thus the learning process of item is the same as entity.
Considering the $j$-th historical items $\mathcal{H}_{d}^j=\{d_1^j,...d_t^j\}$, 
We also use R-GCN and self-attention mechanisms in Eq. (\ref{rgcn}) and (\ref{cur_entity}) to learn the $j$-th session's item representation $\bm{h}_d^j=\textrm{R-GCN}(\mathcal{H}_d^j)$.
%the item matrix $\textrm{V}_d^j$ is also calculated by R-GCN, and $\textrm{V}_d^j$ is averaged like historical entity to derive $\bm{h}_d^j=\textrm{R-GCN}(\textrm{V}_d^j)$.
% 【】这个式子其实不太好，R-GCN最好指eq1，eq2是self-attention。要不就直接
% 【】这个式子中直接给R-GCN那个，然后说明一下V是由R-GCN基于item间的图上得到的aggregation后的矩阵
\begin{comment}
\begin{equation}
    \begin{aligned}
    \bm{h}_i^j &= \mathcal{F}(\bm{\mu}_i^h(\textrm{V}_i^j)^{\textrm{T}}), \\
    \bm{\mu}_i^h &= \textrm{Softmax}(\bm{b}_i^h\textrm{Tanh}(W_i^h\textrm{V}_i^j)).
    \end{aligned}
\end{equation}
\end{comment}
% 正文中解释p_d。
Since the current item is unknown, we use the combination of current words and entities $\bm{p}_d^c$ to represent the current user intention, which is defined as:
%Then the similarity between historical and current items is required to quantify the importance of $\bm{h}_d^j$. However, the current item is unknown~(the aim of our model is to predict the current item). Thus we use the predicted current item representation instead:
\begin{equation}
\label{predicted_consuming}
    \begin{aligned}
    %\bm{r}_i^h &= \sum_{j=1}^{T-1}\varphi(\bm{h}_i^j,\tilde{\bm{p}}_i^c)\ \bm{h}_i^j,\\
    \bm{p}^c_d &= g(\bm{r}_w^c, \bm{r}_e^c) %\tilde{\bm{p}}_i^c (\bm{r}_w^c, \bm{r}_e^c)
    = \tau\cdot \bm{r}_w^c + (1 - \tau)\cdot \bm{r}_e^c,\\
    \tau &= \sigma(W_g\textrm{Concat}(\bm{r}_w^c, \bm{r}_e^c)),
    \end{aligned}
\end{equation}
where $\sigma$ is the Sigmoid activation function.
%Here we choose $\bm{p}_d^c$, as current words and entities are powerful source for modeling user's current consuming preferences.
Finally, the historical item representation is $\bm{r}_d^h=\textrm{Agg}(\bm{p}_d^c, \bm{h}_d^1,...,\bm{h}_d^{T-1})$ like Eq. (\ref{his_entity}), which directly reflect the historical consuming preferences of users.

%As the user mentioned words and entities reflect the user consuming preferences, we leverage the weighted sum of the $\bm{r}_w^c$ and $\bm{r}_e^c$ to model users and provide proper items to users.

We have also tried LSTM or GRU for the historical entity and item learner in the sequential manner. However, the computational cost is high and there is no improvement.
A possible reason is that CRS is different from traditional sequential recommendations, as the current session contains the user's main intention, and is of vital importance. Thus, for the entity and item views that contain concrete structural knowledge, considering the similarity between historical and current sessions is more reasonable than temporal factors,
%would better make use of historical features as supplements, 
and outperform direct sequential models.

\subsection{Multi-View Preference Mapper}
\label{multi_view_align}

\textls[-18]{Now we have the following representations: (1) current entities $\bm{r}_e^c$, (2) current words $\bm{r}_w^c$, (3) historical entities $\bm{r}_e^h$, (4) historical words $\bm{r}_w^h$ and (5) historical items $\bm{r}_i^h$. In this section, we design several self-supervised objectives to learn the intrinsic correlations between different views.}
%Now the users are modeled from three views: word-level semantic view, entity-level knowledge view and item-level consuming view. Considering the historical and current dialogue sessions, we have the following representations: 1). current entities $\bm{r}_e^c$, 2). current words $\bm{r}_w^c$, 3). historical entities $\bm{r}_e^h$, 4). historical words $\bm{r}_w^h$ and 5). historical items $\bm{r}_i^h$. As all the views reflect user preferences, we would like to learn the intrinsic correlations between different views. Thus we design the following auxiliary objectives to model the correlations and learn better representations.
%In this section, we would like to learn the intrinsic correlations between different views. For example, suppose that a user mentioned the word ``scary'' and preferred the horror film ``The Shining'', while another user is interested in ``romantic'' stories such as ``Titanic''. Obviously, ``scary'' and ``The Shining'' are more related as they come from the same user, while ``scary'' and ``Titanic'' are irrelevant. Thus we design the following pre-training objectives to model the correlations and learn better representations.
Inspired by contrastive learning \cite{oord2018representation}, for two view representations $\bm{v}_1$ and $\bm{v}_2$ of a user, we suppose that $\bm{v}_1$ and $\bm{v}_2$ should be more similar than other users'. Given a batch of samples $\mathcal{B}$, we minimize the self-supervised loss to align the two views as follows:
% 【】这里符号体系的问题：前面命名是r_e^c，这里却使用v作为view representation
% 【】引用填对比学习
% 把phi写上去
\begin{equation}
\label{multi_view_mapper}
    \begin{aligned}
    \mathcal{L}_a(\bm{v}_1, \bm{v}_2) &= \sum_{u\in\mathcal{B}}(1-\textrm{sim}(\bm{v}_1^u, \bm{v}_2^{u}))^2 + \lambda_a \sum_{u,u'\in\mathcal{B}}(\textrm{sim}(\bm{v}_1^u, \bm{v}_2^{u'}))^2,
    %\\
    %\phi_{u, u'} &= \textrm{sim}(\bm{v}_1^u, \bm{v}_2^{u'}),
    \end{aligned}
\end{equation}
% 【】\phi没必要，可以直接把sim写到公式
where $\textrm{sim}(\cdot,\cdot)$ is the cosine similarity function that measures the correlation between two views. Here, $u'$ represents the negative users, which are all other users of batch $\mathcal{B}$ except for $u$.

%Specifically, we have three alignment tasks: 1). the current entities and current words~($v_1=\bm{r}_w^c$, $v_2=\bm{r}_e^c$); 2). the historical words and historical entities~($v_1=\bm{r}_w^h$, $v_2=\bm{r}_e^h$); 3). the historical items and the combination of historical words and entities~($v_1=\bm{r}_d^h$, $v_2=\bm{p}_d^h$, $\bm{p}_d^h $ refers to Eq.~\ref{predicted_consuming}).
Specifically, we have three alignment tasks: (1) $\bm{v}_1=\bm{r}_w^c$, $\bm{v}_2=\bm{r}_e^c$; (2) $\bm{v}_1=\bm{r}_w^h$, $\bm{v}_2=\bm{r}_e^h$; (3) $\bm{v}_1=\bm{r}_d^h$, $\bm{v}_2=\bm{p}_d^h$, $\bm{p}_d^h$ is the combination of historical words and entities~(refers to Eq. (\ref{predicted_consuming})).
%By minimizing the alignment loss, the intrinsic correlations between views are injected into the representations~(the weights of the word/entity encoders are trained by $\mathcal{L}_a$), and thus the users are learned better.
Here we do not align the views between current and historical information,
since the user's current intention is not always consistent with historical information in CRS. Instead, we have considered the correlations between historical and current information via current-aware aggregations and temporal factors in Eq. (\ref{his_entity}), (\ref{eq.his_word}) and (\ref{predicted_consuming}).
%As the correlation between historical and current entities has already been injected into $\bm{r}_e^h$ according to Eq.~\ref{his_entity}~($\varphi$ models the history-current correlation), here we do not align the views between historical entities and current entities.
%用DIN构建历史表示，本身就highlight了当前-历史表示，所以就没有再拿进来，有点重复。

% user-CF，用户兴趣会变化，要在用户*时间的维度找相似用户，找到的用户更能匹配。
% 当前行为少的时候作为补充，
% 不同时间段用户功效不同，考虑时间因素。

\subsection{Temporal Look-Alike User Selector}
\label{temp_look_alike_u_sel}

It is widely verified that users having similar historical behaviors are likely to share the same preferences \cite{zhao2010user}. This could be an effective supplement to the user modeling, especially when we can only learn very little from the historical and current sessions for cold-start~\cite{zhu2021learning,pan2019warm,zhu2021transfer,zhu2022personalized} users.
%Thus for each user $u$, we search the look-alike users from the whole user set $\mathcal{U}$ according to the historical representations.
%Then the user $u$ is enhanced by his look-alike user's current behaviors.
However, the user preferences often evolve over dialogues dynamically~\cite{zhou2019deep}. Thus, simply learning one representation for each user is not proper. Here we not only search the look-alike users from the user set $\mathcal{U}$, but also consider the fine-grained user preference slices at every time points for all users.
%Then the user $u$ is enhanced by his look-alike user's current behaviors.
%For example, given the user A who is interested in horror films, and the user B has 10 dialogue sessions. If B also likes horror films in the first session, the behaviors in the second session would enhance the representation of A. However, in the next session, B asks for some ``romantic'' movies instead, and from this session, B is not the look-alike users of A.

Based on the observations, we design a new temporal look-alike user selection, considering user preference evolution in CRS. Concretely, we decompose users into multiple recommendation turns. Suppose that a user $u'$ has $K$ recommendation turns totally (if there are $T$ historical sessions and each session has $t$ recommendation turns, we have $K=t*T$ turns in total). In each recommendation turn, the user may have distinct preferences. Thus, for the word view, at turn $k$, we first calculate the corresponding historical word representation $\bm{r}_w^h(u'_k)$ and current word representation $\bm{r}_w^c(u'_k)$ for user $u'$ at time $k$ (noted as $u'_k$). Then the look-alike user's enhanced representation of user $u$ from user $u'$ is learned as:
% 【】这里之前word和entity写反了
\begin{equation}
\label{look_alike_entity}
    %\delta(\textrm{sim}(u, u'))\ \bm{r}_w^c(u')
    \bm{r}_w^l(u, u')= \sum_{k=1}^K \delta(\textrm{sim}(\bm{r}_w^h(u), \bm{r}_w^h(u'_k)))\ \bm{r}_w^c(u'_k),
\end{equation}
% 【】这里l_e(u,u)像一个函数而不是向量
% l写成向量，l_e^{u, u'}，表达u’对u影响力
\textls[-24]{where $\delta(x) = \textrm{max}(0, x-\delta_w)$ is a clip function to avoid too much noise, and $\delta_w$ is the threshold.
%Besides, $\textrm{sim}(\cdot,\cdot)$ is the same as Eq.~\ref{multi_view_mapper}.
$\bm{r}_w^l(u, u')$ is viewed as the contribution of $u'$ on user $u$'s current word modeling, and $\bm{r}_w^l(u, u')$ equals $0$ when $\textrm{sim}(\bm{r}_w^h(u), \bm{r}_w^h(u'_k))$ is smaller than $\delta_w$ for all time points.
For the entity view, the formalization of $\bm{r}_e^l(u, u')$ is the same as Eq. (\ref{look_alike_entity}).
%For the look-alike users on word view, the definition of $l_w(u, u')$ is the same with Eq.~\ref{look_alike_entity}.
$l_w(u, u')$ and $l_e(u, u')$ will be used as look-alike user supplements for $u$.}

%\textls[-12]{During training, the representations $\bm{r}_e^h$ of users in training sets are recomputed for each epoch. Fortunately, this does not increase too much computation costs, and it will be discussed in Appendix~\ref{app:training}.}

To calculate the look-alike users, for each epoch, we recompute the historical representations $\bm{r}_w^h(u)$ and $\bm{r}_e^h(u)$ for all users at all time points. During training, the similarity function in Eq. (\ref{look_alike_entity}) is only used to distinguish look-alike users. For efficiency, we set a gradient block to $\textrm{sim}(\cdot,\cdot)$ to avoid too much computation.
%and only the loss of $\bm{r}_w^c(u')$ is back-propagated to train the model.
%Thus, only calculating the similarity between users does not cost too much time.

%Here we user word view as a example, the entity view look-alike users is the same. Besides, for convenience, we omit subscript $k$ in the following. The details are discussed in the next session.

%Another difference better other methods is that we use the representations of user mentioned words to learn look-alike users, rather that fixing similar users before training. Thus, the look-alike users would change during the training process, which provides accurate supplements to user modeling. During training, the representations $\bm{r}_w^h$ of users in training sets are recomputed for each epoch. Fortunately, this does not increase too much computation costs, and it will be discussed in Section~\ref{dicussion}.

% 【】这一段motivation还是不够突出，先改其他这个最后看看怎么加强下
% 【】分散在每一段写，感觉不如最后单拿出一段强调一下
% 每个部分提供了什么？有什么设计？在optimization写。

\subsection{Multi-Aspect User-Centric Modeling}
\label{sec.u_centric_m}

In this section, we introduce the user-centric modeling from multi-aspects: current session, historical sessions, and look-alike users.
%Concretely, as users mention multi-view information in dialogues, we perform multi-aspect learning for each view, and then fuse the views to learn user-centric representations.
%In this section, we introduce the user-centric modeling, which understands users from multi-views: semantic, knowledge and consuming views. For each view, we consider multi-aspects: user current representations $\bm{r}_w^c$ and $\bm{r}_e^c$, user historical behaviors $\bm{r}_w^h$, $\bm{r}_e^h$ and $\bm{r}_i^h$, and look-alike users, which performs a comprehensive understanding of users.
%Compare with previous methods which only use current representations $\bm{r}_w^c$ and $\bm{r}_e^c$, our UCCR incorporates the historical behaviors and look-alike users as supplements to user modeling and performs a comprehensive understanding of users.
%对于用户的学习，应该考虑三个view，而每个view又应该考虑三个aspects

\subsubsection{Multi-Aspect Entity View Modeling}
\label{knowledge_view_learning}
% 在subsubsection中，写成entity/word/item views
% 用了3个信息，写Eq.10。为了balance三个信息源，把三部分写出来alpha。

\begin{comment}
The word-level semantic view representation of user $u\in\mathcal{U}$ consists of three parts:
\begin{equation}\label{word_rep}
    \bm{r}_w(u) = \bm{r}_w^c(u) + \alpha_h \bm{r}_w^h(u) + \alpha_s \sum_{u'\in\mathcal{U}} \delta(\textrm{sim}(u, u'))\ \bm{r}_w^c(u'),
\end{equation}
where $\mathcal{U}$ is all the users from training set, and $\bm{r}_w^c(u')$ denotes the current word representations of user $u'$. Next, we introduce the components of $\bm{r}_w(u)$ in details.
\end{comment}

We first introduce the multi-aspect entity view modeling for user $u$ to get the final entity-view representation $\bm{r}_e(u)$ via information from three aspects:
\begin{equation}\label{entity_rep}
    \bm{r}_e(u) = \bm{r}_e^c(u) + \alpha_h \bm{r}_e^h(u) + \alpha_s \sum_{u'\in\mathcal{U}} \bm{r}_e^l(u, u'),
\end{equation}
where $\alpha_h$ and $\alpha_s$ is to regulate the amount of information learned from the historical and look-alike aspects, respectively.

For the historical aspect, the coefficient $\alpha_h$ aims to balance the proportion of historical information incorporated into current session. Thus it is learned by the combination of historical and current representations:
\begin{equation}
\label{his_aspect_word}
    \alpha_h = \mathcal{G}(\textrm{Concat}(\bm{r}_e^c(u), \bm{r}_e^h(u))) \ / \ \tau_e,
\end{equation}
where $\mathcal{G}$ is a feed-forward network with Sigmoid activation function, and $\tau_e$ controls the value of $\alpha_h$.
For the temporal look-alike users representation $\bm{r}_e^l(u,u')$, we also incorporate the coefficient $\alpha_s$. According to the performance on the validation set, we empirically set $\alpha_s$ to 1 for simplification.

\subsubsection{Multi-Aspect Word View Modeling}
\label{word_rep_section}

The word view representation also consists of three aspects like entity view.
%Like the semantic view learning, the entity-level knowledge view representation of user $u\in\mathcal{U}$ also consists of three parts: current entity, historical entity and look-alike users enhanced representations.
Similar with Eq.~\ref{entity_rep}, $\bm{r}_w(u)$ is calculated as:
\begin{equation}\label{word_rep}
    \bm{r}_w(u) = \bm{r}_w^c(u) + \beta_h \bm{r}_w^h(u) + \beta_s \sum_{u'\in\mathcal{U}} \bm{r}_w^l(u, u'),
\end{equation}
and the details could refer to Section~\ref{knowledge_view_learning}.

\subsubsection{Multi-Aspect Item View Modeling}

As user's current preferred items are not available, the current aspect of item view is unknown and the recommender task is to predict the current item. Thus, the look-alike aspect is also not applicable to enhance item view user modeling~(not applicable for current item). Thus the consuming view is calculated via historical item representation $\bm{r}_d^h(u)$ as:
%The difference between the item-level consuming view and the above is that the user's current preferred items are not available~(the task of CRS is to predict the user's current consuming items). Thus we do not know the current consuming representations $\bm{r}_d^c$. Moreover, the look-alike user's current consuming representations are also not available to enhance the users. Based on the observations above, the consuming view representation of $u$ is defined as:
\begin{equation}
\begin{aligned}
    &\bm{r}_d(u)=\gamma_h \bm{r}_d^h(u),\\
    &\gamma_h= \delta(\textrm{sim}(\bm{p}_d^h,\ \bm{p}_d^c)),
\end{aligned}
\end{equation}
where $\gamma_h$ also balances the proportion of historical and current information. As the current consuming item is unknown, we use the combination of mentioned words and entities instead, and $\gamma_h$ is calculated by the combination of $\bm{p}_d^h$~(historical user intent) and $\bm{p}_d^c$~(current user intent). To be noticed, the defined of $\bm{p}_d^h$ and $\bm{p}_d^c$ is the same as Eq. (\ref{predicted_consuming}), $\delta(\cdot)$ and $\textrm{sim}(\cdot,\cdot)$ is the same as Eq. (\ref{look_alike_entity}).

% 【】eq.11和12合成一行

\subsubsection{Multi-Aspect Multi-View Fusion}

With the multi-view representations learning, the user representation is calculated as:
\begin{equation}
    \bm{r}(u) = g(\bm{r}_w(u), \ \bm{r}_e(u)) + \bm{r}_d(u),
\end{equation}
where $g(\cdot,\cdot)$ is the combination of words and entities like Eq. (\ref{predicted_consuming}).
%$\bm{r}(u)$ models users by multi-view behaviors learning~(including word-level semantic view, entity-level knowledge view and item-level consuming view) from multi-aspects~(including current behavior, historical behavior and look-alike users), which performs a comprehensive understanding of users.

% 【】为什么是r(u)而不是r_u呢？明明是一个向量。下一章转置那个就看起来很奇怪

\subsection{Optimization}

The learned user representation $\bm{r}(u)$ is leveraged to both provide high-quality recommendations and generate utterances.

\subsubsection{Recommendation Objective}

The probability of recommending item $d_i$ to user $u$ is calculated by the user representation $\bm{r}(u)$:
\begin{equation}
    \bm{p}_{rec}(u, d_i) = \textrm{Softmax}(\bm{r(u)}^{\top} \cdot \bm{d}_i),% 这里[d]是什么意思啊
\end{equation}
where $\bm{d}_i$ is the representation of item $d_i$. Then we adopt cross-entropy loss to train the recommendation model:
\begin{equation}\label{rec_loss}
    \mathcal{L}_{rec} = -\ \sum_{u\in\mathcal{U}}\sum_{i=1}^{N_u}\log \bm{p}_{rec}(u, d_i) + \lambda_{CL}\sum_{(v_1,v_2)}\mathcal{L}_a(v_1, v_2),
\end{equation}
where $\mathcal{L}_a(v_1,v_2)$ is the multi-view preference alignment loss.

% 【】按道理来说，这个对比的loss不应该加到recommendation，而是加在全局吧
% lambda_CL有点奇怪也没有解释和实验

\subsubsection{Dialogue Generation Objective}

For dialogue generation, we adopt the standard seq2seq framework~\cite{sutskever2014sequence} following~\cite{chen2019towards,zhou2020improving}. Concretely, Transformer~\cite{vaswani2017attention} is used as the base model for encoder and decoder, which consists of several multi-head attention layers and fully connected feed-forward layers.

Given the input utterances, the encoder first extracts the semantic feature and the decoder outputs a representation $\bm{q}$ for token generation. To incorporate the user preferences into token generation, we use $\bm{r}(u)$ as a bias feature:
\begin{equation}
    \bm{p}_{dial}(y_t|y_1,...,y_{t-1}) = \textrm{Softmax}(W^G\bm{q}+\mathcal{M}(\bm{r}(u))[y_t],
\end{equation}
where $\mathcal{M}$ is a linear transformation which guarantees the dimension of $\mathcal{M}(\cdot)$ equals the vocabulary size. As $\bm{r}(u)$ is injected into $\bm{p}_{dial}$, the generated utterances satisfy the user needs better. Then the dialogue module is trained with the cross-entropy loss:
\begin{equation}\label{vocab_bias}
    \mathcal{L}_{dial} = -\sum_{u\in\mathcal{U}}\sum_{t=2}^{N_t}\log(\bm{p}_{dial}(y_t|y_1,...,y_{t-1})).
\end{equation}

\subsection{Motivations and Discussions on UCCR}
In this work, we systematically consider the current session, historical sessions, and look-alike users in CRS. Compared with previous CRS methods, which only use the current session features or the ``historical'' sentences/turns in the current session, our UCCR performs a comprehensive understanding of users in CRS. Here we give detailed discussions on all model designs.

\noindent
\textbf{Current Session Learner.} It is an important source to learn user preferences and capture user intentions in CRS. We simply follow the previous works~\cite{chen2019towards,zhou2020improving} to encode the current session, which is not the focused contribution of our user-centric modeling in UCCR. 

\noindent
\textbf{Historical Session Learner.} The historical sessions are completely different from the dialogue/conversation history used in previous CRS works~\cite{chen2019towards,zhou2020improving,zhou2020towards}, as their ``historical'' information is actually the historical turns of the current dialogue session. The usages of historical sessions in CRS are also largely different from the historical user-item interactions~\cite{chang2021sequential,wang2021sequential} in traditional recommendations, since the recommendation in CRS is strongly constrained by the current user intentions, while traditional recommendations are not.

%Based on the intrinsic difference between CRS and common recommendation, directly adding historical and look-alike features is not proper. Thus the usage of multi-aspects features is non-common and our UCCR is carefully designed. For historical sessions, to guarantee that the useful part of historical data is captured, the historical session learner~(Sec.~\ref{his_sess_learner}) considers the similarity between historical and current sessions for entities and the temporal factor for words. Moreover, we design several self-supervised objects~(Sec.~\ref{multi_view_align}) to aligns multiple views and learn the intrinsic correlations between them. For look-alike users selectors~(Sec.~\ref{temp_look_alike_u_sel}), as the user preferences are changing dynamically, it selects look-alike users for each timestamp, rather than static similar users. Finally, when fusing the multiple aspect features~(Sec.~\ref{u_centric_m}), we also controls the amount of useful historical and look-alike features to avoid ruining the real user preferences. All these techniques are novel and suitable to the CRS scenario. The overview illustration of our model is shown in Fig~\ref{network}.

Thus, the main goal of historical sessions learner is: \textit{learning current-related and regular information from historical sessions, without impeding the current session information.} To achieve this goal, we \textit{model the relation between historical and current session.} For the structural historical entities and items, which are concrete objects, we directly use the current representations to filter the useful information. For the historical words, which contain general semantic knowledge, we simply use the temporal factor to weigh them. Both of two designs could effectively extract current-related information from historical sessions. This historical information is beneficial especially when the current session contains little information.

\noindent
\textbf{Temporal Look-alike User Selector.} The main goal of temporal look-alike user selector is: \textit{leveraging the similar users' current features to enhance the user modeling}. Here the similar users are calculated by the historical features. Moreover, in CRS, user preferences often change over dialogue sessions. Thus we take the user interest evolution into consideration, selecting accurate look-alike features from each time point. When both the current and historical information is little, the look-alike feature is a useful supplement.

\noindent
\textbf{Multi-Aspect User-Centric Modeling.} To jointly add historical and look-alike features without confusing the current preference, we also consider the balance between the current and historical user preferences via our multi-aspect user-centric modeling. Thus, all three aspects could jointly provide a comprehensive user modeling.

\begin{table*}[!th]
	\small
	\centering
	\setlength{\tabcolsep}{3pt}
	\caption{The recommendation results. The marker * indicates that the improvement is statistically significant compared with the best baseline (t-test with p-value < 0.05).} \label{rec_results}
	\begin{tabular}{@{}c|cccccc|cccccc@{}}
		\toprule
		Dataset & \multicolumn{5}{c}{TG-ReDial} & & \multicolumn{6}{c}{ReDial}\\
		\midrule
		\textbf{Method} & HR@10 & HR@50 & MRR@10 & MRR@50 & NDCG@10 & NDCG@50 & HR@10 & HR@50 & MRR@10 & MRR@50 & NDCG@10 & NDCG@50 \\
		\midrule
		
		SASRec & 0.0048 & 0.0170 & 0.0011 & 0.0016 & 0.0019 & 0.0046 & 0.0418 & 0.1598 & 0.0385 & 0.0407 & 0.0473 & 0.0712 \\
		Text CNN & 0.0052 & 0.0188 & 0.0015 & 0.0022 & 0.0029 & 0.0058 & 0.0733 & 0.1810 & 0.0438 & 0.0482 & 0.0576 & 0.0808 \\
		Bert & 0.0098 & 0.0356 & 0.0027 & 0.0040 & 0.0051 & 0.0101 & 0.1499 & 0.2937 & 0.0683 & 0.0761 & 0.0813 & 0.1167 \\
		ReDial & 0.0102 & 0.0370 & 0.0028 & 0.0041 & 0.0053 & 0.0107 & 0.1733 & 0.3359 & 0.0779 & 0.0841 & 0.0969 & 0.1351 \\
		KBRD & 0.0141 & 0.0481 & 0.0045 & 0.0063 & 0.0072 & 0.0143 & 0.1827 & 0.3688 & 0.0784 & 0.0855 & 0.1004 & 0.1428 \\
		TG-ReDial & 0.0168 & 0.0513 & 0.0061 & 0.0080 & 0.0088 & 0.0161 & 0.1893 & 0.3801 & 0.0801 & 0.0883 & 0.1032 & 0.1477 \\
		KGSF & 0.0175 & 0.0543 & 0.0073 & 0.0088 & 0.0096 & 0.0175 & 0.2006 & 0.4034 & 0.0837 & 0.0932 & 0.1110 & 0.1556 \\
		KECRS & 0.0113 & 0.0394 & 0.0033 & 0.0042 & 0.0057 & 0.0111 & 0.1772 & 0.3423 & 0.0780 & 0.0851 & 0.0983 & 0.1391 \\
		RevCore & 0.0191 & 0.0581 & 0.0077 & 0.0093 & 0.0105 & 0.0189 & 0.2058 & 0.4088 & 0.0850 & 0.0946 & 0.1132 & 0.1583 \\
		\midrule
		UCCR w/o En & 0.0167 & 0.0506 & 0.0071 & 0.0085 & 0.0092 & 0.0165 & 0.1976 & 0.3885 & 0.0812 & 0.0908 & 0.1084 & 0.1502 \\
		UCCR w/o Wo & 0.0207 & 0.0592 & 0.0080 & 0.0095 & 0.0114 & 0.0196 & 0.2106 & 0.4196 & 0.0865 & 0.0959 & 0.1168 & 0.1613 \\
		UCCR w/o It & 0.0211 & 0.0626 & 0.0082 & 0.0098 & 0.0116 & 0.0201 & 0.2146 & 0.4193 & 0.0865 & 0.0966 & 0.1173 & 0.1619 \\
		%Ours+2 & 3.21 & 8.99 & 0.0117 & 0.0142 & 0.0164 & 0.0289 \\
		UCCR & $\textbf{0.0232}^*$ & $\textbf{0.0664}^*$ & $\textbf{0.0088}^*$ & $\textbf{0.0107}^*$ & $\textbf{0.0122}^*$ & $\textbf{0.0214}^*$ & $\textbf{0.2161}^*$ & $\textbf{0.4258}^*$ & $\textbf{0.0883}^*$ & $\textbf{0.0981}^*$ & $\textbf{0.1182}^*$ & $\textbf{0.1642}^*$ \\
		\bottomrule
	\end{tabular}
	%\vspace{-2mm}
\end{table*}

\section{Experiment}

To validate the superiority of UCCR, we conduct extensive evaluations to answer the following research questions:
\textbf{RQ1}: How does our UCCR perform on the recommendation and dialogue generation tasks compared with the state-of-the-art baselines?
%How does the recommendation and dialogue generation performances of UCCR compared with the state-of-the-art baselines in CRS?
\textbf{RQ2}: What are the benefits of UCCR in cold-start scenarios?
\textbf{RQ3}: How do different components of UCCR benefit its performance, i.e., different views and different aspects?
\textbf{RQ4}: How do different hyper-parameter settings impact UCCR?
%\textbf{RQ5}: How does UCCR models users? Are the learned representations reasonable?

%实验问题：
%1、baseline加上SOTA的说法
%2、对话部分评测方法；为什么对话部分有提升
%3、数据划分实验是？用户序列兴趣会变，考虑user preference、考虑用户历史时序。
%不同时间段数量不一定可比
%个人时间线、全局时间线。第一次考虑历史，因此要这样划，follow一些传统工作划法。。发现这个现象，证明时间穿越。在一个小地方说一下。
%4、table6的实验想说明什么？提升幅度和主实验也是差不多的呀？
%证明在冷启动场景下提升更大
%word少的，entity少的，word和entity都少
%真实世界中长尾的数据更多，用户都是信息量比较少的。

%user 向量 session向量，相同的user的向量会更近，随时间变化

\subsection{Experimental Settings}

\subsubsection{Datasets}

\textls[-0]{We conduct experiments on two widely-used public datasets collected from the real-world platforms, including both Chinese~(TG-ReDial~\cite{zhou2020towards}) and English~(ReDial~\cite{li2018towards}) languages. ReDial contains 10,006 dialogues consisting of 504 users related to 51,699 movies. TG-ReDial contains 10,000 dialogues consisting of 1,482 users related to 33,834 movies.
% 【】数据集给引用
Since we highlight the historical dialogue sessions in CRS, two datasets are split according to the \textbf{chronological order}. We randomly choose several users, using their last several dialogue sessions as our validation and test sets. The remaining sessions are the train set. Here we choose the last two sessions for TG-ReDial and the last four sessions for ReDial to guarantee that the whole train/validation/test sample ratio is about 8:1:1.
%The reason is that the number of users in ReDial is less than TG-ReDial, thus we choose more users for ReDial to guarantees that the whole sample ratio is about 8:1:1.
In ReDial, some users have less than four dialogue sessions (i.e., they have no historical session information in the train set). These users are also used for evaluating UCCR in cold-start scenarios.}
%As we consider the users and historical dialogue sessions in CRS, the datasets are split according to \textit{chronological} orders. The ratio of samples in train, validation and test sets is about 8:1:1. More details could refer to appendix~\ref{dataset}.
%To evaluate the effectiveness of our proposed method, we conduct experiments on two real-world datasets
%\footnote{They can be downloaded at \url{https://github.com/RUCAIBox/CRSLab}}, including both Chinese and English languages.

%They are split into training, validation and test sets using a ratio of 8:1:1, which is the same as
%the one adopted by other baseline methods~\cite{chen2019towards,zhou2020improving,zhou2020towards}.
%other baseline methods~\cite{chen2019towards,zhou2020improving,zhou2020towards}.

%\noindent
\subsubsection{Baselines}
\textls[-12]{Following~\cite{zhou2020towards}, we evaluate the superiority of our UCCR by considering the following nine representative baselines: (1) \textbf{\textit{SASRec}}~\cite{kang2018self} only leverages user historical items for recommendation. (2) \textbf{\textit{Text CNN}}~\cite{kim2014convolutional} encodes utterances in the current session to learn user preferences by CNN-based model. (3) \textbf{\textit{BERT}}~\cite{kenton2019bert} is a pre-training~\cite{zeng2021knowledge} model that encodes current utterances for recommendation. (4) \textbf{\textit{ReDial}}~\cite{li2018towards} is a CRS method which adopts an auto-encoder framework. (5) \textbf{\textit{KBRD}}~\cite{chen2019towards} adopts the external knowledge graph DBpedia for user mentioned entities in current dialogue session to enhance the user representations. (6) \textbf{\textit{TG-ReDial}}~\cite{zhou2020towards} presents the task of topic-guided conversational recommendation, which incorporates topic threads to control the dialogue state transitions. (7) \textbf{\textit{KGSF}}~\cite{zhou2020improving} incorporates both the semantic and KG information for modeling user preferences. They use mutual information maximization to align representations of words and entities from current dialogue session. (8) \textbf{\textit{KECRS}}~\cite{zhang2021kecrs} proposes bag-of-entity with a high-quality KG to better capture user preferences. (9) \textbf{\textit{RevCore}}~\cite{lu2021revcore} incorporates the user reviews on movies to enhance CRS models.}

Among baselines, \textit{SASRec}, \textit{Text CNN} and \textit{Bert} are classical recommendation methods, and \textit{ReDial}, \textit{KBRD}, \textit{TG-ReDial}, \textit{KGSF}, \textit{KECRS} and \textit{RevCore} are CRS methods. All these methods only consider the current dialogue session.
%We do not compare with \textit{TG-ReDial} in dialogue evaluation, since it adopts extra pre-train models for generation, which is not a fair comparison with other methods.
Besides, we do not compare~\cite{lu2021revcore} since it needs external user review information.
For fair comparisons, we implement all the baselines and UCCR by the open-source toolkit CRSLab~\cite{zhou2021crslab}. The hyper-parameter settings of baselines follow the default settings on CRSLab, which reaches the best performances. Note that the results of CRS methods on TG-ReDial are slightly lower than the public results\footnote{\url{https://github.com/RUCAIBox/CRSLab}}, as we split the two datasets by \textbf{chronological} order, while previous work simply randomly split the samples, which omits the user's historical information.

%\noindent
\subsubsection{Evaluation Metrics}

\textls[-12]{The recommender module and dialogue generation module are evaluated separately. For the recommender part, we want to know whether UCCR models the user preferences and provides high-quality recommendations accurately. Thus, we adopt HR@$k$, MRR@$k$ and NDCG@$k$ for evaluation ($k = 10, 50$)\footnote{Following \url{https://github.com/RUCAIBox/CRSLab}.}. For the dialogue module, we consider both automatic and human evaluations. In the automatic evaluations, we adopt BLEU-2,3~\cite{papineni2002bleu} and perplexity (PPL) for testing the accuracy and fluency of generations, and Distinct $n$-gram~\cite{li2016diversity,zhou2020towards} ($n = 2, 3, 4$) for the diversity. In the human evaluations, three annotators are invited to score the \textit{Fluency} and \textit{Informativeness} of the generated responses. The range of scores is from 0 to 2, and the scores of three annotators are averaged.}

%\noindent
\subsubsection{Implementation Details}

The dimensions of embeddings are set to 128 and 300 for recommendation and dialogue respectively. The number of layers is set to 1 for both R-GCN and GCN considering effectiveness and efficiency. The hyper-parameters of both historical entities learner $\lambda_e$~(Eq. (\ref{his_entity})) and historical items learner $\lambda_i$ are set to 0.1, and $\lambda_a$~(Eq. (\ref{multi_view_mapper})) in multi-view preference mapper equals 0.1. For the historical aspect of user-centric modeling, both $\tau_w$~(Eq. (\ref{his_aspect_word})) for words and $\tau_e$ for entities are set to 6. For the look-alike aspect, both $\delta_w$~(Eq. (\ref{word_rep})) and $\delta_e$~(Eq. (\ref{entity_rep})) are set to 0.85. Finally, the weight of multi-view preferences mapper loss (in Eq. (\ref{rec_loss})) is 0.025. All of them are selected by grid search on the validation set. For training, we adopt the Adam optimizer~\cite{kingma2014adam} with a learning rate of 0.001, where the batch size is set as 128. The epochs of preference mapper training are 3, and we train the model 25 epochs for both recommendation and dialogue tasks. For baselines, hyper-parameter settings follow their own implementations, which reaches the best performances and guarantees fair comparisons.

\begin{table*}[!th]
	\small
	\centering
	\setlength{\tabcolsep}{2pt}
	\caption{Results on dialogue generation. Flu. and Inf. stand for Fluency and Informativeness, respectively. The marker * indicates that the improvement is statistically significant compared with the best baseline (t-test with p-value< 0.05).} \label{dial_results}
	\begin{tabular}{@{}c|cccccccc|cccccccc@{}}
		\toprule
		Dataset & \multicolumn{7}{c}{TG-ReDial} & & \multicolumn{7}{c}{ReDial}\\
		\midrule
		\textbf{Method} & Bleu-2 & Bleu-3 & Dist-2 & Dist-3 & Dist-4 & PPL & Flu. & Inf. & Bleu-2 & Bleu-3 & Dist-2 & Dist-3 & Dist-4 & PPL & Flu. & Inf.\\
		\midrule

		ReDial & 0.0409 & 0.0102 & 0.2672 & 0.5288 & 0.8012 & 55.71 & 0.71 & 0.75 & 0.0217 & 0.0078 & 0.0689 & 0.2697 & 0.4638 & 56.21 & 0.73 & 0.91 \\
		KBRD & 0.0423 & 0.0119 & 0.3482 & 0.6911 & 0.9972 & 53.08 & 0.83 & 0.88 & 0.0238 & 0.0088 & 0.0712 & 0.2883 & 0.4893 & 54.89 & 0.82 & 1.00  \\
		KGSF & 0.0461 & 0.0135 & 0.4447 & 1.0450 & 1.5792 & 51.27 & 1.01 & 1.09 & 0.0249 & 0.0091 & 0.0756 & 0.3024 & 0.5177 & 54.75 & 0.95 & 1.14 \\
		KECRS & 0.0332 & 0.0078 & 0.1893 & 0.3799 & 0.6531 & 58.97 & 0.63 & 0.64 & 0.0133 & 0.0051 & 0.0473 & 0.1532 & 0.3188 & 59.35 & 0.59 & 0.71 \\
		RevCore & 0.0467 & 0.0136 & 0.4513 & 1.0932 & 1.6631 & 51.03 & 1.06 & 1.11 & 0.0252 & 0.0098 & 0.0769 & 0.3065 & 0.5283 & 54.43 & 0.98 & 1.15 \\
		\midrule
		UCCR w/o En & 0.0465 & 0.0138 & 0.4349 & 1.0289 & 1.5543 & 51.33 & 1.02 & 1.08 & 0.0245 & 0.0089 & 0.0729 & 0.3001 & 0.5082 & 54.95 & 0.96 & 1.12 \\
		UCCR w/o Wo & 0.0478 & 0.0141 & 0.5093 & 1.2239 & 1.8583 & 50.68 & 1.07 & 1.14 & 0.0253 & 0.0097 & 0.0801 & 0.3195 & 0.5493 & 54.01 & 1.00 & 1.18 \\
		UCCR w/o It & 0.0481 & 0.0142 & 0.5217 & 1.2589 & 1.9122 & 50.34 & 1.08 & 1.16 & 0.0255 & 0.0103 & 0.0815 & 0.3255 & 0.5561 & 53.56 & 1.03 & 1.18 \\
		UCCR & $\textbf{0.0494}^*$ & $\textbf{0.0145}^*$ & $\textbf{0.5365}^*$ & $\textbf{1.2783}^*$ & $\textbf{1.9376}^*$ & $\textbf{50.21}^*$ & $\textbf{1.13}^*$ & $\textbf{1.18}^*$ & $\textbf{0.0257}^*$ & $\textbf{0.0106}^*$ & $\textbf{0.0818}^*$ & $\textbf{0.3289}^*$ & $\textbf{0.5635}^*$ & $\textbf{53.24}^*$ & $\textbf{1.06}^*$ & $\textbf{1.22}^*$ \\
		\bottomrule
	\end{tabular}
	%\vspace{-2mm}
\end{table*}

% 【】这里为什么没有加TG-ReDial的实验结果
% 【】TG-ReDial的命名是否可以没有歧义

\subsection{Overall Performance (RQ1)}

\subsubsection{Recommendation}

The recommendation module's results on two datasets are shown in Table~\ref{rec_results}. Based on the results, we can see that our UCCR significantly outperforms all the baselines by a large margin on both two datasets, which verifies that UCCR could successfully capture multi-aspect multi-view user preferences and achieve the SOTA performances under the user-centric manner.
We analyze the effectiveness of our UCCR compared with different baselines as follows:

First, UCCR outperforms the six CRS methods, this shows the effectiveness of the user-centric modeling, which can understand users from multiple aspects. Although the current session is very important in CRS, the historical session and look-alike user aspects are also essential supplements to model users' diverse preferences accurately, especially when there is little information in the current session.
Moreover, our multi-view preference mapper also provides additional training for extracting intrinsic correlations between different views, which helps to build better user representations.

Then, our UCCR significantly outperforms the non-CRS method SASRec. SASRec is a competitive sequential recommendation~(SR) method which only uses historical items to learn user preferences, and it performs badly in CRS.
The reason is that SASRec ignores the core features in CRS, i.e. current session, and only depends on historical items for user modeling. In contrast, UCCR models the correlations between historical and current sessions properly in historical session learner, which brings in additional user preferences from historical sessions related to the current user intention.

Finally, our UCCR beats the non-CRS methods Text CNN and BERT, for they directly model user preferences from the contextual utterances. We can find that useful information in natural language is sparse and hard to extract. Hence, considering multi-view information from multiple aspects in UCCR is also essential.

\subsubsection{Dialogue Generation.}

We also evaluate UCCR on the dialogue generation task. Here we do not compare with \textit{TG-ReDial} in dialogue evaluation, since it adopts an extra pre-train model GPT-2 for generation, which is not fair comparing with other methods.
%GPT-2~\cite{radford2019language}, which is not a fair comparison with our Transformer-based generator.

The results are shown in Table~\ref{dial_results}, and we can see that: (1) UCCR generates more fluent, diverse, and informative utterances from both automatic and human evaluations perspectives, compared with the baseline methods. The main reason is that UCCR provides better user representations by considering the historical features and look-alike users, and they serve as vocabulary bias~(in Eq.~\ref{vocab_bias}) to generate proper tokens. Thus, better user representations also improve the generation quality. (2) Besides, compared with ReDial, we can see that the external knowledge graph of entities and semantic similarity information also contribute to better generations.

%From the results, first, we can see that KGSF performs best among baselines, which indicates that external knowledge graph of entities and semantic similarity information also contribute to better generations.
%Then, our UCCR generates more fluent, diverse and informative utterances from both automatic and human evaluations perspectives.
%The main reason is that UCCR provides better user representations, and they serve as vocabulary bias~(refer to Eq.~\ref{vocab_bias}) to generate proper tokens. Thus, the better user representations also improves the generation quality.

\subsection{Results on Cold-Start Scenarios (RQ2)}

In this section, we further evaluate UCCR in the cold-start scenarios from two perspectives: (1) the current information is limited, and (2) the historical information is limited, which are practical in CRS.

\subsubsection{Cold-Start Current Information}

%For previous CRS methods, they mainly focus on the current dialogue session~(i.e. current entities and words).
Cold-start issues are common and critical in real-world CRS. Nearly 55\% recommendations occur when the user mentioned entities of the current session are no more than 2 in ReDial.
%especially at the first recommendation of a dialogue session.
Thus we simulate this scenario by considering the recommendations with few current entities, i.e., users only mention 0,1,2,3 entities in the current session.

\begin{table}[!th]
	\small
	\centering
	\vspace{-0mm}
	\setlength{\tabcolsep}{3pt}
	\caption{Results of cold-start scenarios on ReDial with different number of user's current entities.}
	%Incorporating historical information could improve the performances.}
	\vspace{-2mm}
	\label{num_cur_entities}
	\begin{tabular}{@{}c|ccccccc@{}}
		\toprule
		\textbf{\#Entity} & \textbf{Method} & H@10 & H@50 & M@10 & M@50 & N@10 & N@50 \\
		\midrule
		
		%Ours+2 & 3.21 & 8.99 & 0.0117 & 0.0142 & 0.0164 & 0.0289 \\
		\multirow{2}{*}{0} & RevCore & 10.23 & 26.31 & 0.0317 & 0.0409 & 0.0483 & 0.0799 \\
		& UCCR & \textbf{11.61} & \textbf{28.36} & \textbf{0.0384} & \textbf{0.0471} & \textbf{0.0574} & \textbf{0.0906} \\
		\midrule
		\multirow{2}{*}{1} & RevCore & 23.88 & 41.76 & 0.1094 & 0.1186 & 0.1377 & 0.1764 \\
		& UCCR & \textbf{24.69} & \textbf{43.93} & \textbf{0.1153} & \textbf{0.1231} & \textbf{0.1409} & \textbf{0.1830} \\
		\midrule
		\multirow{2}{*}{2} & RevCore & 22.65 & 41.92 & 0.0939 & 0.1045 & 0.1271 & 0.1693 \\
		& UCCR & \textbf{23.44} & \textbf{42.12} & \textbf{0.0996} & \textbf{0.1084} & \textbf{0.1313} & \textbf{0.1725} \\
		\midrule
		\multirow{2}{*}{3} & RevCore & 23.15 & 44.69 & 0.0859 & 0.0967 & 0.1202 & 0.1684 \\
		& UCCR & \textbf{23.41} & \textbf{44.95} & \textbf{0.0886} & \textbf{0.0987} & \textbf{0.1214} & \textbf{0.1703} \\
		\midrule
		\multirow{2}{*}{$\geq$ 6} & RevCore & 18.63 & 40.77 & 0.0789 & 0.0898 & 0.1048 & 0.1562 \\
		& UCCR & \textbf{19.28} & \textbf{41.64} & \textbf{0.0829} & \textbf{0.0942} & \textbf{0.1116} & \textbf{0.1617} \\
		\bottomrule
	\end{tabular}
	\vspace{-1.6mm}
\end{table}

% 【】需要显著性实验

\textls[-16]{The results are shown in Table~\ref{num_cur_entities}. We can see that:
(1) UCCR outperforms RevCore for all numbers of current entities. Especially, when there is no current entity, the performance of RevCore is poor as RevCore only leverages current features for user modeling. While UCCR outperforms RevCore significantly as the user-centric modeling with multi-aspect information. It reconfirms that the historical features and look-alike users are powerful supplements to the user preferences modeling, and our UCCR learns multi-aspects user features appropriately.
(2) As the number of current entities increases, the gap between RevCore and UCCR gradually narrowed, while UCCR consistently beats RevCore.
(3) Moreover, we also show that our historical and look-alike features are not only useful in the beginning stage of the dialogue~(lack of current entities), but also in the latter stage~(plenty of current entities). In the last row of Table~\ref{num_cur_entities}, we show the situation that current entities are no less than 6 (about 10\% of the whole test set), where UCCR also outperforms RevCore.}

%1. 为什么要引入历史？如果不引入历史，像传统方法独立考虑每篇对话，建模对话，会怎样？每篇对话的结果会收到时间这个因素的影响，下表这个实验说明了时间靠前对话，准确率更高。因此前人的方法是不合适的，数据计划分也不合理。因此要考虑到用户的历史信息，同时我们这样划分数据集也是最合理的。
%此处也可以改成，为何要向我们这样划分数据集？说明以前的划分方式不合理，同时考虑历史信息是有必要的。

%2. 引入历史后的好处？（可能需要把相似用户part去掉）a. 对每篇对话开始时，即当前words/entities很少时，历史数据可以起到有效辅助。即考虑当前每篇对话中第1、2、3次推荐的效果。（即当前entity / word很少时，引入历史带来的提升）（应该用ReDial数据集）b.(这个做不了！测试集本来就历史很多) 当用户历史很少时，我们的方法也很稳定，能够充分用户历史数据。

\subsubsection{Cold-Start Historical Information.}
\label{cold_start_history}

\begin{figure}[htbp]
	\centering
	\subfigure[MRR\&NDCG results for new users.]{
		\begin{minipage}[t]{0.63\linewidth}
			\centering
			\includegraphics[width=0.95\textwidth]{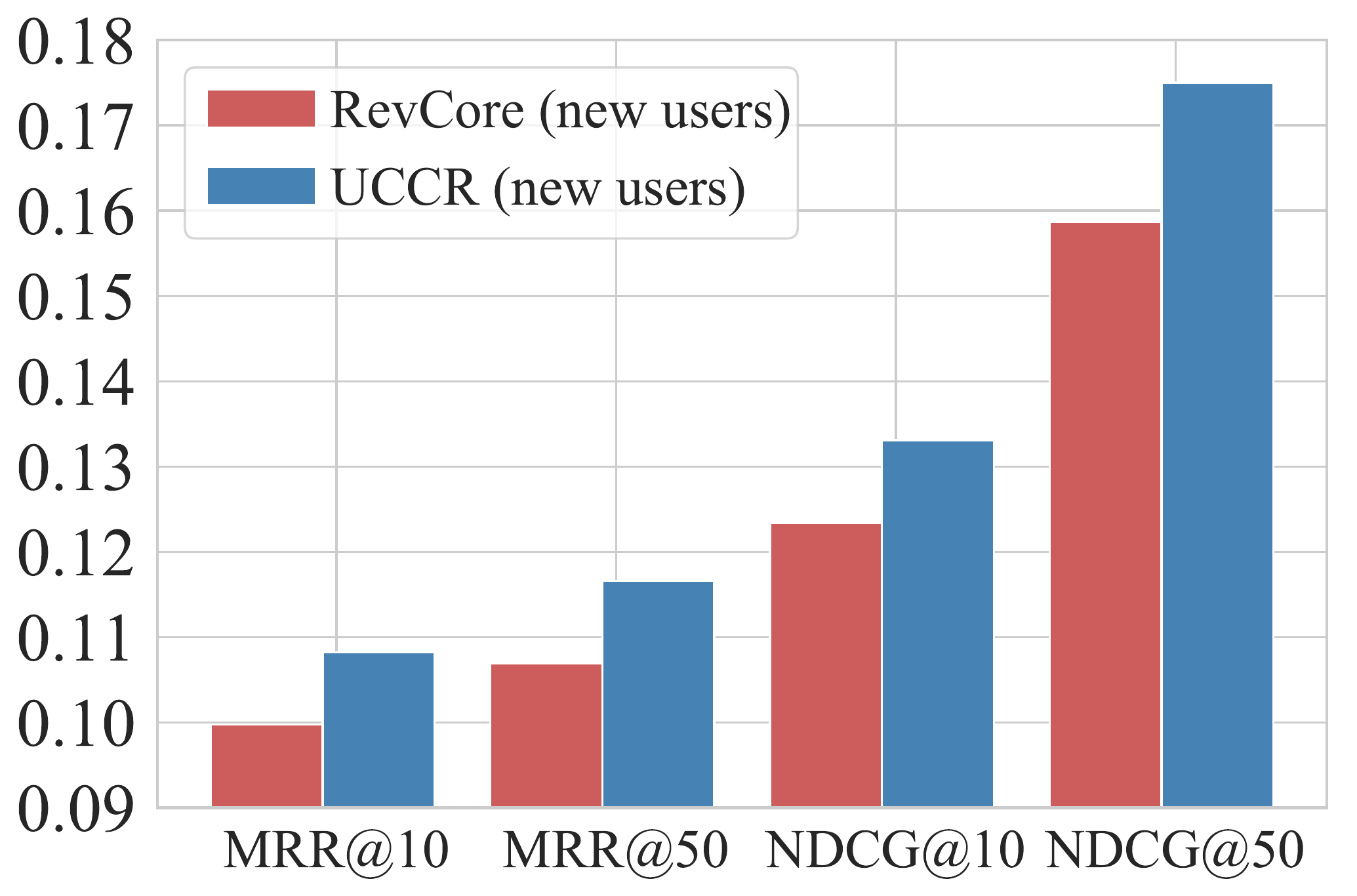}
			%\caption{fig1}
		\end{minipage}%
	}%
	\subfigure[HR results for new users.]{
		\begin{minipage}[t]{0.355\linewidth}
			\centering
			\includegraphics[width=0.95\textwidth]{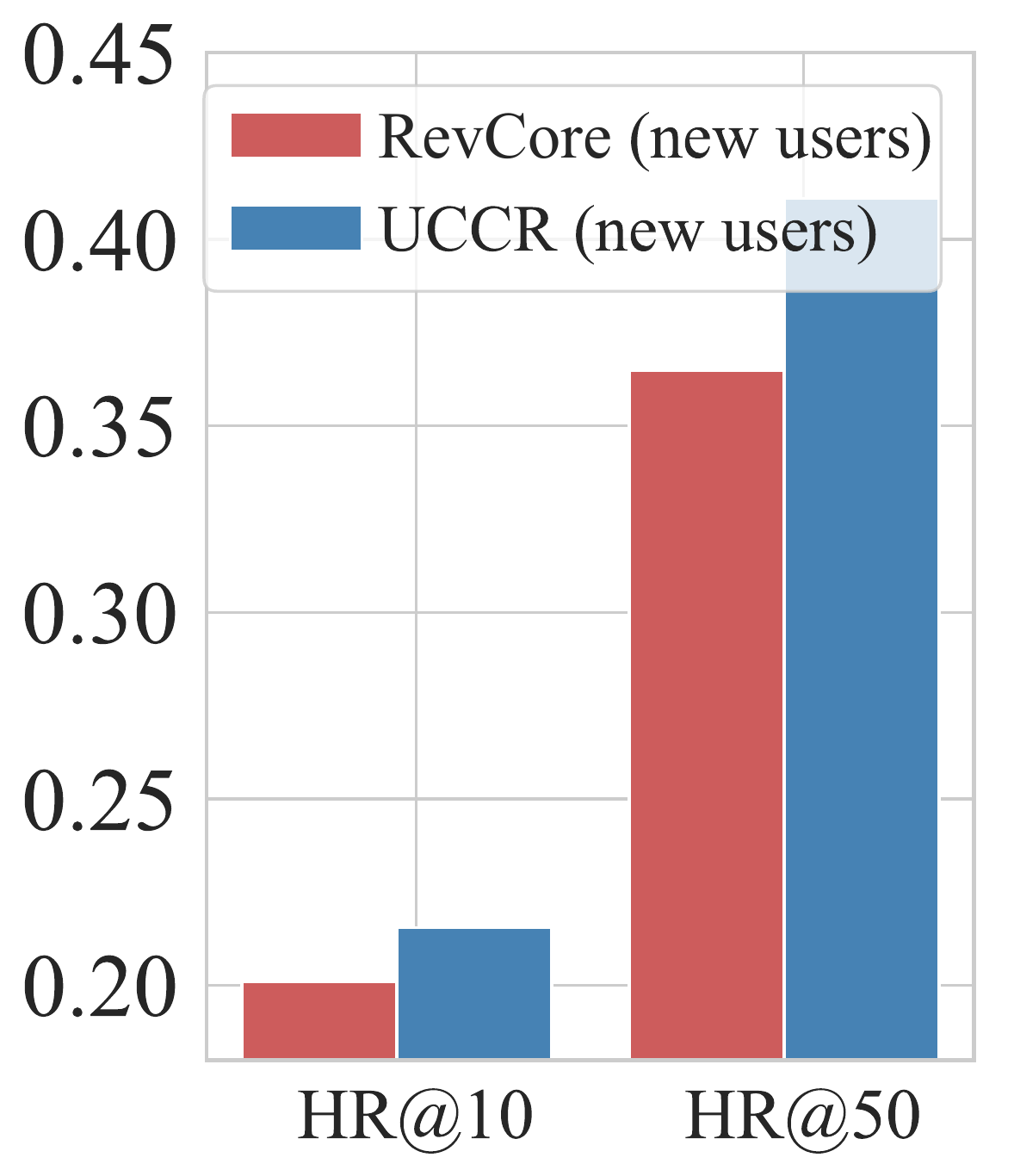}
			%\caption{fig2}
		\end{minipage}%
	}%
	
	\centering
	\subfigure[MRR\&NDCG results for old users.]{
		\begin{minipage}[t]{0.62\linewidth}
			\centering
			\includegraphics[width=0.95\textwidth]{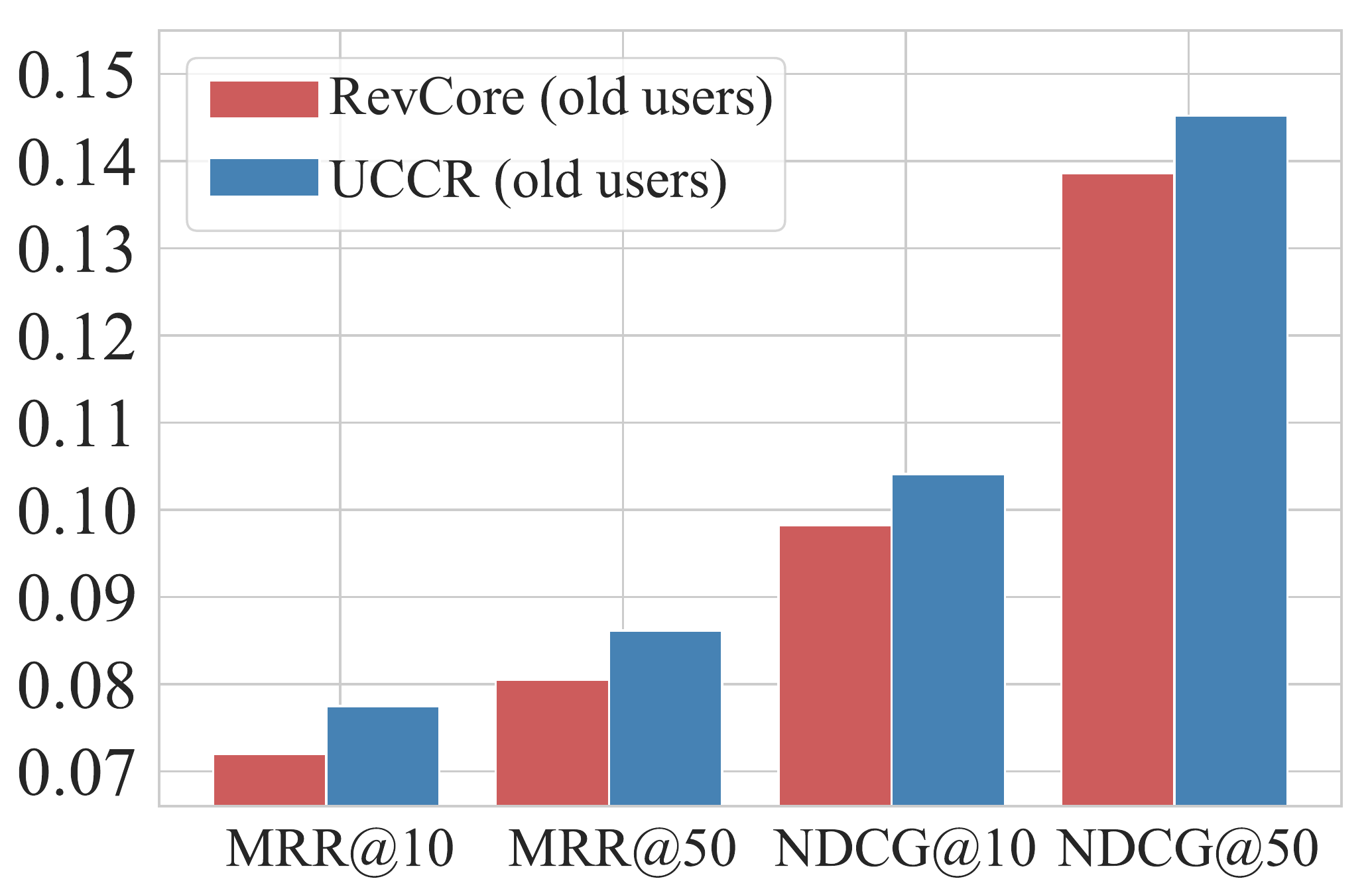}
			%\caption{fig1}
		\end{minipage}%
	}%
	\subfigure[HR results for old users.]{
		\begin{minipage}[t]{0.365\linewidth}
			\centering
			\includegraphics[width=0.95\textwidth]{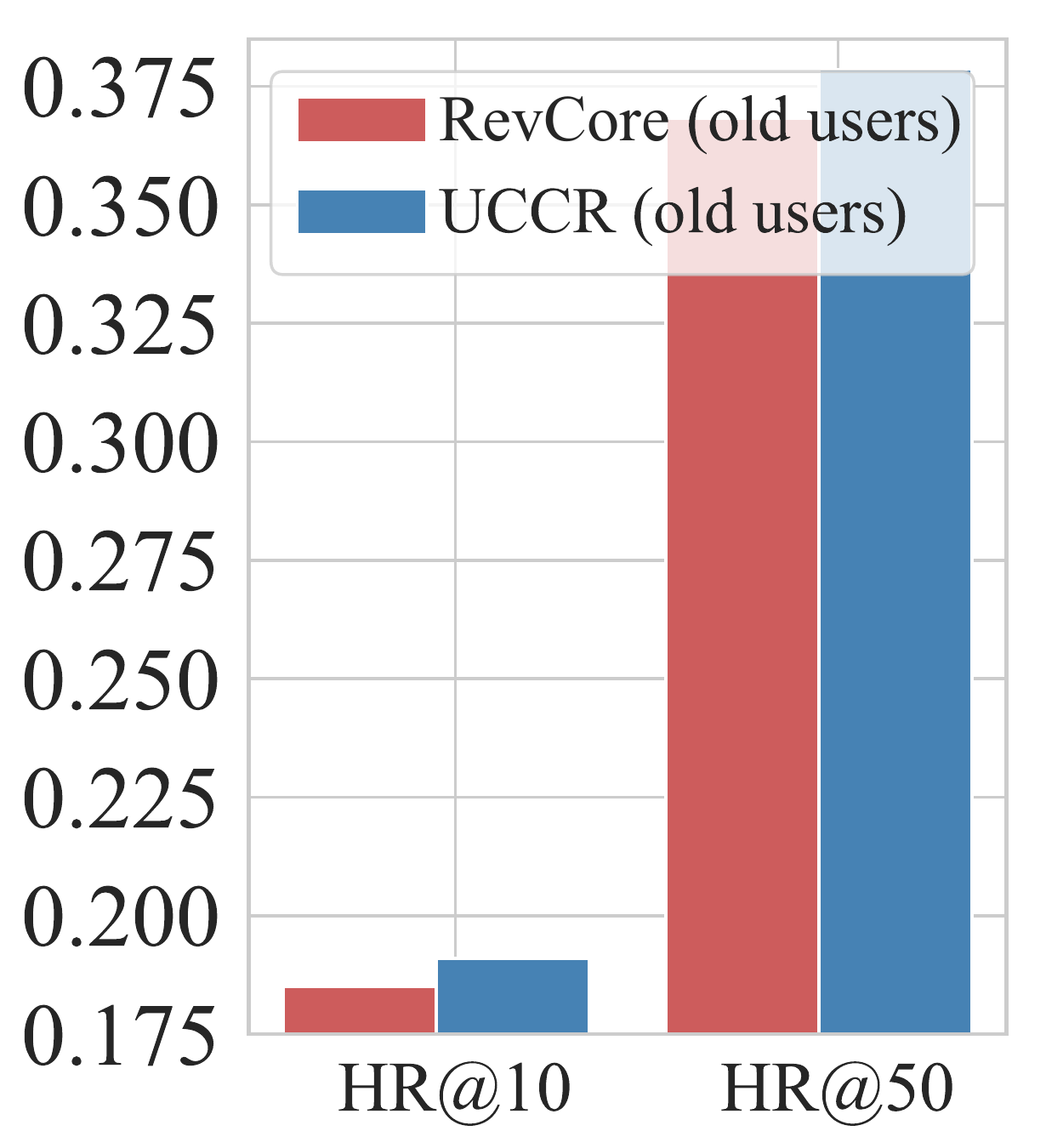}
			%\caption{fig2}
		\end{minipage}%
	}%
	\centering
	\vspace{-2.15mm}
	\caption{\textls[-16]{The results for cold-start historical sessions scenario.}}
	\vspace{-2.15mm}
	\label{new_old_users}
\end{figure}

In UCCR, we consider user historical features for user modeling, while some users have no historical dialogue sessions in practice. Hence, to evaluate UCCR with little historical information, we split users into new and old users according to whether they have historical sessions.
Fig. \ref{new_old_users} presents the results on ReDial. We can find that: UCCR performs well for both old and new users, especially for new users. The improvement ratios of new and old users are about 10.3\% V.S. 4.8\% on NDCG@50 and 12.6\% V.S. 2.8\% on HR@50. Although there are no historical sessions for new users, UCCR still learns better user preferences by multi-aspects user modeling (especially from the look-alike user aspect). Thus, our UCCR is applicable to all users.

\begin{table}[!th]
	\small
	\centering
	\setlength{\tabcolsep}{3pt}
	\vspace{0mm}
	\caption{Ablation study for three aspects.} \label{more_ablation}
	\vspace{0mm}
	\begin{tabular}{l|cccccc}
		\toprule
		 & H@10 & H@50 & M@10 & M@50 & N@10 & N@50 \\
		\midrule
		UCCR w/o En & 1.67 & 5.06 & 0.0071 & 0.0085 & 0.0092 & 0.0165 \\
		%Ours+2 & 3.21 & 8.99 & 0.0117 & 0.0142 & 0.0164 & 0.0289 \\
		\midrule
		\quad + Current & 1.95 & 6.12 & 0.0076 & 0.0093 & 0.0103 & 0.0192 \\
		\quad + Historical & 2.14 & 6.33 & 0.0082 & 0.0101 & 0.0114 & 0.0204 \\
		\quad + Look-alike & 2.32 & 6.64 & 0.0088 & 0.0107 & 0.0122 & 0.0214 \\
		\bottomrule
	\end{tabular}
	\vspace{-1mm}
	%\vspace{-2mm}
\end{table}

\subsection{Ablation Study (RQ3)}
\label{ablation}

To evaluate the effectiveness of each part of UCCR, we conduct the ablation study for both multiple views and aspects.

First, we show the impacts of each view in Table~\ref{rec_results} and Table~\ref{dial_results}. Specifically, we report the results with different views turned off. Here En, Wo, and It refer to the entity, word, and item views. From the results, we can observe that all the three views contribute to the main model, as the performance decreases with any of the views removed. Another observation is that the entity view is of vital importance and the result w/o En drops largely, even worse than KGSF in Table~\ref{rec_results}. A possible reason is that entities contain structural knowledge, and are more correlated with recommended movies.

%Besides, note that the results drop slightly when the historical items missing, which further proves that current features are most important in CRS and historical features are supplements.

Then, we show the effectiveness of each aspect of user modeling in Table~\ref{more_ablation}. As entity view impacts the performance most, we take entity view for example, and add each aspect~(i.e. current, historical, and look-alike) into UCCR w/o En. We further conduct t-test~(p-value$<$0.05) for each aspect to confirm the statistical significance.
With no doubt, when adding the current features, it improves mostly, which indicates that current features are of vital importance in CRS. When adding other two aspects, the results also rise, which shows that they are powerful supplements to user modeling in CRS. Besides, note that the results drop slightly when the historical items missing in Table~\ref{rec_results}, which further proves that current features are most important in CRS.

%Here we take entity view for example, and adding each aspect~(i.e. current, historical and look-alike) into UCCR w/o En. With no doubt, when adding the current features, it improves mostly, which indicates that current features are of vital importance in CRS.

% 【】可以说明一下简单的设计会使得结果下降，对模型complexity的回答

\subsection{Parameter Sensitive Analysis (RQ4)}

In this section, we investigate the sensitivity of hyper-parameters in historical aspect $\tau_e$ and look-alike aspect $\delta_e$~(in Sec.~\ref{knowledge_view_learning}).

\begin{figure}[htbp]
\vspace{-2mm}
\centering
\subfigure[$\tau_e$ for historical aspect.]{
\begin{minipage}[t]{0.48\linewidth}
\centering
\includegraphics[width=0.8\textwidth]{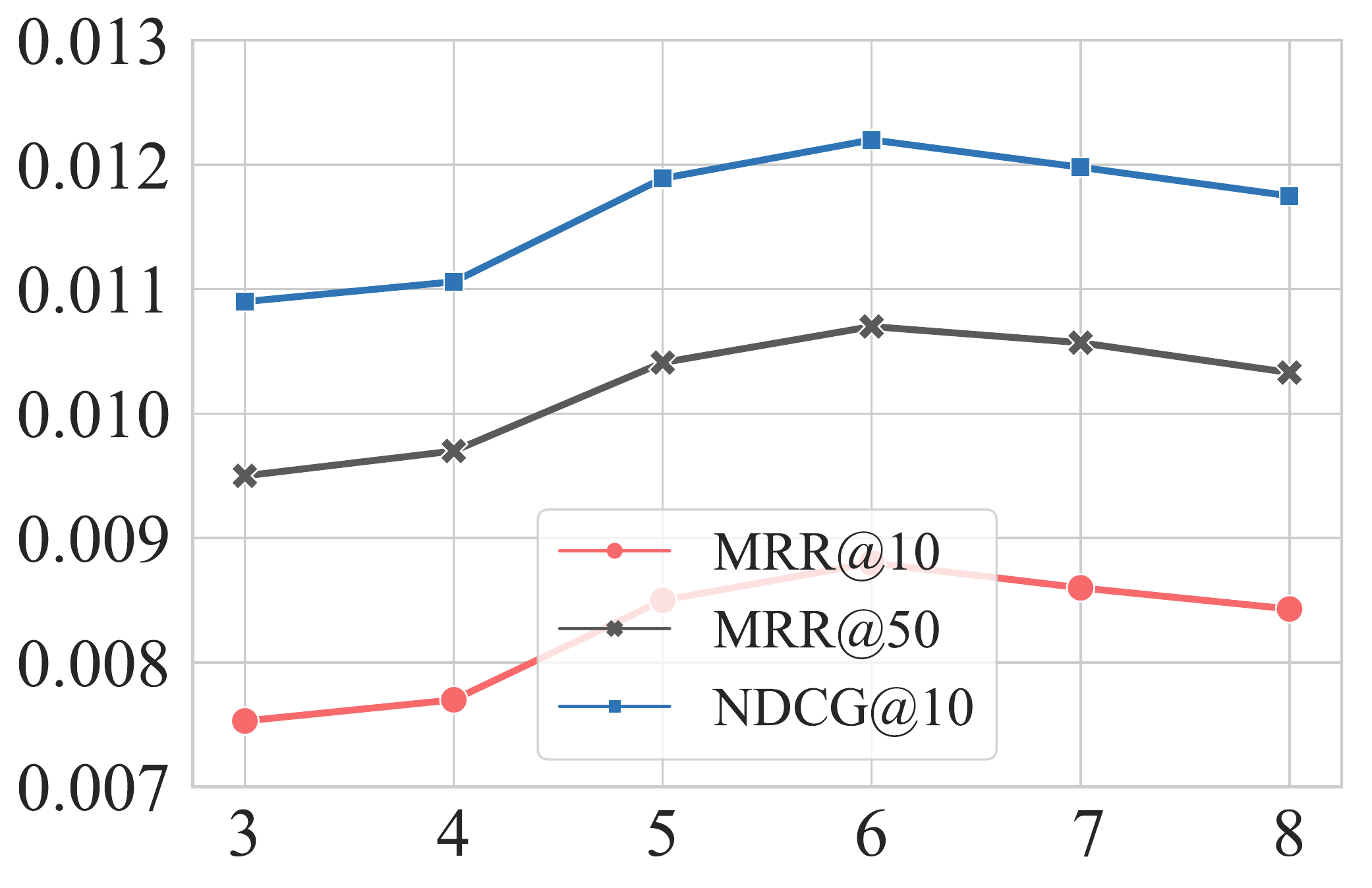}
%\caption{fig2}
\end{minipage}%
}%
\subfigure[$\delta_e$ for look-alike aspect.]{
\begin{minipage}[t]{0.48\linewidth}
\centering
\includegraphics[width=0.8\textwidth]{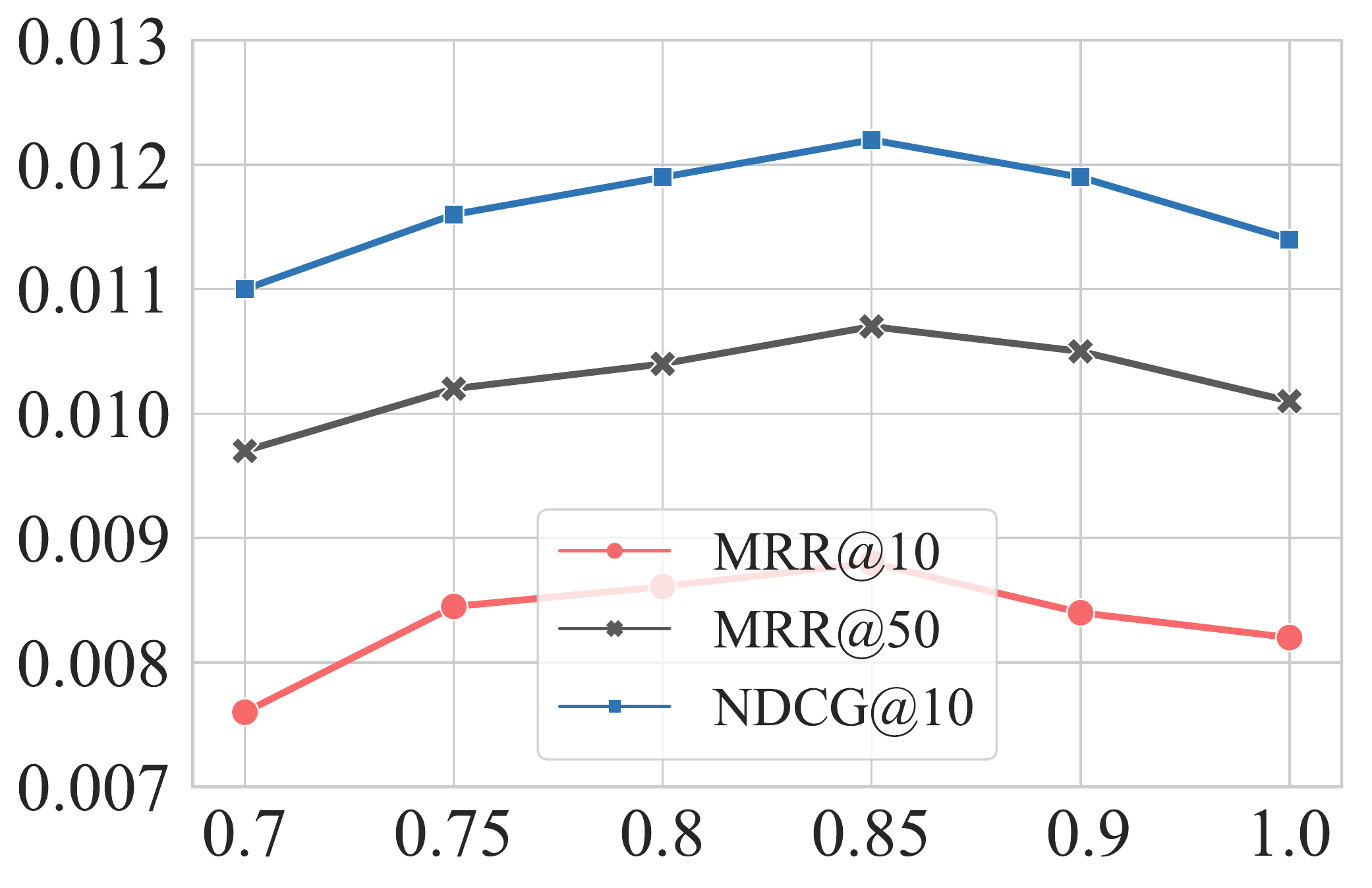}
%\caption{fig1}
\end{minipage}%
}%
\vspace{-2mm}
\caption{Hyper-parameters sensitive analysis for historical entities and look-alike users on TG-Redial.}
\vspace{-2mm}
\label{para_sensi}
\end{figure}

For each hyper-parameter, we search the value in empirical intervals, and six representative values are reported in Fig.~\ref{para_sensi}. In general, the results first rise and then decline with the increase of the hyper-parameters values. And our UCCR performs well in a wide range of $\tau_e$ and $\delta_e$ values. In a word, both historical and look-alike aspects contribute to the recommendation performances, and we should tune them in fine-grained.

%在ReDial数据集中，有些用户的总对话篇数很少，所以有些用户仅仅在测试集出现，训练集中不出现。所以new user定义为训练集中没出现过的新用户；old user定义为训练集中出现的用户。同时为了公平对比，我们取用户current entity小于3的场景。
%结果显示无论新用户或者老用户，UCCR都有明显提升。特别的，UCCR对于新用户提升更大，因此UCCR对用户建模更准确，在冷启动场景下也能取得很好的效果。

%case study：用户相似度受到历史、当前信息共同影响。总的来说，历史较为相似的用户，相似度会更高（user 1，2；user 3，4），因此历史相似的用户，更可能share同样的current preference；当前信息也会极大影响用户相似性，如（user 3，4相似度要低于user 1，2）。

%3. 为什么要考虑相似用户？统计分析，历史提到words、entities相互重叠的用户，他们喜欢的items也更多的重合？（或是share一阶邻居？）

%4. 引入相似用户后模型的表现？（可能需要把用户历史part去掉）针对不同类型的用户（按照相似用户数量来划分），效果都有了显著提升。

%后续分析
%1. 参数敏感度分析
%2. ablation 整体的和局部的

%case study 找几个用户具体看看。。 历史比较相似的用户，user current rep也比较相似，也即user的当前兴趣比较相似。因此通过历史找相似用户，是比较合理的。
%那如何体现user当前表示和历史表示的关系？能否把单个user当前-历史和相似用户之间的当前-当前画到同一张图里？

\section{Related Work}

Conversational Recommender System~(CRS)~\cite{chen2019towards,lesi2020interactive,xu2021adapting} focus on capturing user preferences through dialogues and provide high-quality items. One category is attribute-based conversational recommendation systems~\cite{christakopoulou2016towards,xu2021adapting,sun2018conversational,li2021seamlessly,zhou2020leveraging,fu2021hoops,ren2021learning,xie2021comparison}, which only care about providing high-quality recommendations and do not pay much attention to utterances generation. They converse with users via the pre-defined questions, which have pre-defined slots and patterns. 
%They mainly ask about items or attributes to capture user preferences, and do not systematically consider the multiple features of users. For example, ~\cite{zhou2020leveraging} does not use the historical session. ~\cite{li2021seamlessly} captures user preferences by multi-armed bandit approach, and does not consider the personalized user historical features.

\textls[-24]{Recently, several works~\cite{li2018towards,chen2019towards, zhou2020improving,zhou2020towards,liu2020towards} focus to build end-to-end conversational recommender systems. They aim to provide high-quality recommendations and generate fluent utterances, simultaneously. ~\cite{li2018towards} releases a CRS dataset coined ReDial and proposes an HRED-based~\cite{serban2017hierarchical} model. Then, the following works mainly focus on the current user features, and pay much attention to natural language understanding: incorporating external knowledge~\cite{chen2019towards,sarkar2020suggest,zhou2020improving}, controlling dialogue policy~\cite{liu2020towards,zhou2020towards}, using user reviews for movies~\cite{lu2021revcore}, designing templates for generations~\cite{liang2021learning}. For external knowledge, ~\cite{chen2019towards} incorporates the entity knowledge graph for representation learning. Then ~\cite{zhou2020improving} incorporates both the entity knowledge (DBpedia) and the semantic similarity of words (ConceptNet). For dialogue policy, ~\cite{liu2020towards} divides a conversation into several goals with goal planning, and ~\cite{zhou2020towards} uses topics to guide dialogues. They care about generating proactive and natural human-like utterances.
For movie reviews, ~\cite{lu2021revcore} collects user reviews on movies to better model user preferences. Finally, ~\cite{liang2021learning} learns templates for utterances generation and does not consider the recommendation task.}

These CRS methods only use the information from the current dialogue session, and omit the historical session and look-alike features. To be noticed, the dialogue/conversation history mentioned in~\cite{chen2019towards,zhou2020improving,zhou2020towards} is not the historical sessions, as their ``historical" information is actually historical turns of the current dialogue session. Moreover, historical sessions are also different from historical user-item interactions in traditional recommendations, as in CRS, the relation between historical and current sessions should be balanced.

%Different from the above approaches, our work mainly focuses on a comprehensive user modeling in end-to-end CRS. Besides the current user features, we incorporate the historical features and look-alike users, which returns to central subjects~(i.e. users) and proposes multi-aspect user modeling. To the best of our knowledge, we are the first to systematically model users from multiple aspects in end-to-end CRS. Moreover, historical sessions are also different from historical user-item interactions in traditional recommendations

\begin{comment}
\textbf{Multi-View Learning in Recommendation} \textls[-12]{There are mainly two kinds of methods for multi-view representation learning: 1). multi-view representation fusion~\cite{he2017neural,zheng2019explore}, which the multi-view data is fused into a single compact vector; 2). multi-view representation alignment~\cite{elkahky2015multi,jiang2015deep}, which aims to model intrinsic correlations among different views by alignment tasks. Recently, several multi-view methods~\cite{wang2020m2grl,zheng2021multi} are designed for graph data representations learning.}
\textbf{User Modelling}
\end{comment}

% 【】需要强调这些工作和我们工作的不同
% 【】一个是和CRS中考虑historical xx的差别：他们没有使用multi-view historical session信息
% 【】一个是和传统推荐的差别，很不一样，简单提一下就行。

\section{Conclusion and Future Work}

\textls[-24]{In this paper, we proposed a novel method UCCR for a comprehensive user modeling in CRS. Different previous methods which only focus on the current session and dialogue understanding, UCCR returns to the central subjects in CRS~(i.e. users). UCCR models users by considering multi-aspect multi-view information from the current session, historical sessions, and look-alike users. The relations between different aspects are further explored to properly leverage historical and look-alike features. Extensive experiments verify the effectiveness of UCCR.}

In the future, we will use more sophisticated methods to better model all aspects. We will also explore the correlations between different aspects to further improve the user understanding in CRS.

\section*{Acknowledgement}
The research work is supported by the National Key Research and Development Program of China under Grant No. 2017YFB1002104. This work is also supported by Alibaba Group through Alibaba Innovative
Research Program and the National Natural Science Foundation of China under Grant (No.61976204, U1811461, U1836206). Xiang Ao is also supported by the Project of Youth Innovation Promotion Association CAS, Beijing Nova Program Z201100006820062.

\bibliographystyle{ACM-Reference-Format}
\bibliography{sample-base}

\end{document}